\documentclass{aa}  

\usepackage{graphicx}
\usepackage{txfonts}
\usepackage{xcolor}
\usepackage{lastpage}
\usepackage{orcidlink}

\newcommand{\oiii}{[O\,\textsc{iii}]}
\newcommand{\nii}{[N\,\textsc{ii}]}
\newcommand{\sii}{[S\,\textsc{ii}]}
\newcommand{\oi}{[O\,\textsc{i}]}

\newcommand{\siii}{[S\,\textsc{iii}]}
\newcommand{\oii}{[O\,\textsc{ii}]}

\newcommand{\hii}{H\,\textsc{ii}}
\newcommand{\ha}{H$\alpha$}
\newcommand{\hb}{H$\beta$}
\newcommand{\niiauroral}{[N\,\textsc{ii}]$\lambda$5755}
\newcommand{\doh}{$\Delta$(O/H)}
\newcommand{\sigmaoh}{$\sigma$(O/H)}
\newcommand{\te}{$T_{\rm e}$}
\newcommand{\logten}{log$_{10}$}
\newcommand{\nee}{$n_{\rm e}$}

\usepackage{color}
\usepackage{natbib,twoopt}
\usepackage[hyphenbreaks]{breakurl}
\usepackage{hyperref}      
\bibpunct{(}{)}{;}{a}{}{,}             
\hypersetup{colorlinks=true,linkcolor=blue,citecolor=blue,filecolor=blue,urlcolor=blue}
\makeatletter
  \newcommandtwoopt{\citeads}[3][][]{\href{http://adsabs.harvard.edu/abs/#3}%
    {\def\hyper@linkstart##1##2{}%
     \let\hyper@linkend\@empty\citealp[#1][#2]{#3}}}
  \newcommandtwoopt{\citepads}[3][][]{\href{http://adsabs.harvard.edu/abs/#3}%
    {\def\hyper@linkstart##1##2{}%
     \let\hyper@linkend\@empty\citep[#1][#2]{#3}}}
  \newcommandtwoopt{\citetads}[3][][]{\href{http://adsabs.harvard.edu/abs/#3}%
    {\def\hyper@linkstart##1##2{}%
     \let\hyper@linkend\@empty\citet[#1][#2]{#3}}}
  \newcommandtwoopt{\citeyearads}[3][][]%
    {\href{http://adsabs.harvard.edu/abs/#3}
    {\def\hyper@linkstart##1##2{}%
     \let\hyper@linkend\@empty\citeyear[#1][#2]{#3}}}
\makeatother

\begin{document}

   \title{
   Temperature based radial metallicity gradients in nearby galaxies 
   }


\newcommand{\OSU}{\label{OSU} Department of Astronomy, The Ohio State University, 140 West 18th Avenue, Columbus, Ohio 43210, USA}

\newcommand{\Alberta}{\label{Alberta} Department of Physics, University of Alberta, Edmonton, AB T6G 2E1, Canada}

\newcommand{\ANU}{\label{ANU} Research School of Astronomy and Astrophysics, Australian National University, Canberra, ACT 2611, Australia}

\newcommand{\IPAC}{\label{IPAC} Caltech-IPAC, 1200 E. California Blvd. Pasadena, CA 91125, USA}

\newcommand{\Carnegie}{\label{Carnegie} Observatories of the Carnegie Institution for Science, 813 Santa Barbara Street, Pasadena, CA 91101, USA}

\newcommand{\CCAPP}{\label{CCAPP} Center for Cosmology and Astroparticle Physics, 191 West Woodruff Avenue, Columbus, OH 43210, USA}

\newcommand{\CfA}{\label{CfA} Harvard-Smithsonian Center for Astrophysics, 60 Garden Street, Cambridge, MA 02138, USA}

\newcommand{\CITEVA}{\label{CITEVA} Centro de Astronomía (CITEVA), Universidad de Antofagasta, Avenida Angamos 601, Antofagasta, Chile}

\newcommand{\CNRS}{\label{CNRS} CNRS, IRAP, 9 Av. du Colonel Roche, BP 44346, F-31028 Toulouse cedex 4, France}

\newcommand{\ESO}{\label{ESO} European Southern Observatory, Karl-Schwarzschild Stra{\ss}e 2, D-85748 Garching bei M\"{u}nchen, Germany}

\newcommand{\ESOChile}{\label{ESOChile} European Southern Observatory, Avenida Alonso de Cordoba 3107, Casilla 19, Santiago 19001, Chile}

\newcommand{\HD}{\label{HD} Astronomisches Rechen-Institut, Zentrum f\"{u}r Astronomie der Universit\"{a}t Heidelberg, M\"{o}nchhofstra\ss e 12-14, D-69120 Heidelberg, Germany}

\newcommand{\ICRAR}{\label{ICRAR} International Centre for Radio Astronomy Research, University of Western Australia, 35 Stirling Highway, Crawley, WA 6009, Australia}

\newcommand{\IRAM}{\label{IRAM} Institut de Radioastronomie Millim\'{e}trique (IRAM), 300 Rue de la Piscine, F-38406 Saint Martin d'H\`{e}res, France}

\newcommand{\ITA}{\label{ITA} Universit\"{a}t Heidelberg, Zentrum f\"{u}r Astronomie, Institut f\"{u}r Theoretische Astrophysik, Albert-Ueberle-Str 2, D-69120 Heidelberg, Germany}

\newcommand{\IWR}{\label{IWR} Universit\"{a}t Heidelberg, Interdisziplin\"{a}res Zentrum f\"{u}r Wissenschaftliches Rechnen, Im Neuenheimer Feld 205, D-69120 Heidelberg, Germany}

\newcommand{\JHU}{\label{JHU} Department of Physics and Astronomy, The Johns Hopkins University, Baltimore, MD 21218, USA}

\newcommand{\Leiden}{\label{Leiden} Leiden Observatory, Leiden University, P.O. Box 9513, 2300 RA Leiden, The Netherlands}

\newcommand{\Maryland}{\label{Maryland} Department of Astronomy, University of Maryland, College Park, MD 20742, USA}

\newcommand{\MPE}{\label{MPE} Max-Planck-Institut f\"{u}r extraterrestrische Physik, Giessenbachstra{\ss}e 1, D-85748 Garching, Germany}

\newcommand{\MPIA}{\label{MPIA} Max-Planck-Institut f\"{u}r Astronomie, K\"{o}nigstuhl 17, D-69117, Heidelberg, Germany}

\newcommand{\Nagoya}{\label{Nagoya} Department of Physics, Nagoya University, Furo-cho, Chikusa-ku, Nagoya, Aichi 464-8602, Japan}

\newcommand{\NRAO}{\label{NRAO} National Radio Astronomy Observatory, 520 Edgemont Road, Charlottesville, VA 22903-2475, USA}

\newcommand{\OAN}{\label{OAN} Observatorio Astron\'{o}mico Nacional (IGN), C/Alfonso XII, 3, E-28014 Madrid, Spain}

\newcommand{\ObsParis}{\label{ObsParis} Sorbonne Universit\'{e}, Observatoire de Paris, Universit\'{e} PSL, CNRS, LERMA, F-75014, Paris, France}

\newcommand{\Princeton}{\label{Princeton} Department of Astrophysical Sciences, Princeton University, Princeton, NJ 08544 USA}

\newcommand{\UToledo}{\label{UToledo} University of Toledo, 2801 W. Bancroft St., Mail Stop 111, Toledo, OH, 43606}

\newcommand{\Toulouse}{\label{Toulouse} Universit\'{e} de Toulouse, UPS-OMP, IRAP, F-31028 Toulouse cedex 4, France}

\newcommand{\UBonn}{\label{UBonn} Argelander-Institut f\"ur Astronomie, Universit\"at Bonn, Auf dem H\"ugel 71, 53121 Bonn, Germany}

\newcommand{\UChile}{\label{UChile} Departamento de Astronom\'{i}a, Universidad de Chile, Camino del Observatorio 1515, Las Condes, Santiago, Chile}

\newcommand{\UConn}{\label{UConn} Department of Physics, University of Connecticut, Storrs, CT, 06269, USA}

\newcommand{\UCSD}{\label{UCSD} Center for Astrophysics and Space Sciences, Department of Physics,  University of California, San Diego, 9500 Gilman Drive, La Jolla, CA 92093, USA}

\newcommand{\UGent}{\label{UGent} Sterrenkundig Observatorium, Universiteit Gent, Krijgslaan 281 S9, B-9000 Gent, Belgium}

\newcommand{\ULyon}{\label{ULyon} Univ Lyon, Univ Lyon 1, ENS de Lyon, CNRS, Centre de Recherche Astrophysique de Lyon UMR5574,\\ F-69230 Saint-Genis-Laval, France}

\newcommand{\UMass}{\label{UMass} University of Massachusetts—Amherst, 710 N. Pleasant Street, Amherst, MA 01003, USA}

\newcommand{\UWyoming}{\label{UWyoming} Department of Physics and Astronomy, University of Wyoming, Laramie, WY 82071, USA}

\newcommand{\LAM}{\label{LAM} Aix Marseille Univ, CNRS, CNES, LAM (Laboratoire d’Astrophysique de Marseille), Marseille, France}

\newcommand{\UHawaii}{\label{UHawaii} Institute for Astronomy, University of Hawaii, 2680 Woodlawn Drive, Honolulu, HI 96822, USA}

\newcommand{\UCM}{\label{UCM} Departamento de F\'{\i}sica de la Tierra y Astrof\'{\i}sica, Universidad Complutense de Madrid, E-28040, Spain}

\newcommand{\IPARC}{\label{IPARC} Instituto de F\'{\i}sica de Part\'{\i}culas y del Cosmos IPARCOS, Facultad de Ciencias F\'{\i}sicas, Universidad Complutense de Madrid, E-28040, Spain}

\newcommand{\STScI}{\label{STScI} Space Telescope Science Institute, 3700 San Martin Drive, Baltimore, MD 21218, USA}

\newcommand{\McMaster}{\label{McMaster} Department of Physics and Astronomy, McMaster University, 1280 Main Street West, Hamilton, ON L8S 4M1, Canada}

\newcommand{\INAF}{\label{INAF} INAF -- Osservatorio Astrofisico di Arcetri, Largo E. Fermi 5, I-50157, Firenze, Italy}

\newcommand{\Sydney}{\label{Sydney} Sydney Institute for Astronomy, School of Physics A28, The University of Sydney, NSW 2006, Australia}

\newcommand{\CITA}{\label{CITA} Canadian Institute for Theoretical Astrophysics (CITA), University of Toronto, 60 St George St, Toronto, ON M5S 3H8, Canada}

\newcommand{\ASIAA}{\label{ASIAA} Institute of Astronomy and Astrophysics, Academia Sinica, No. 1, Sec. 4, Roosevelt Road, Taipei 10617, Taiwan}

\newcommand{\TKU}{\label{TKU} Department of Physics, Tamkang University, No.151, Yingzhuan Rd., Tamsui Dist., New Taipei City 251301, Taiwan}

\newcommand{\PSMA}{\label{PSMA} Penn State Mont Alto, 1 Campus Drive, Mont Alto, PA  17237, USA}

\newcommand{\ILL}{\label{ILL} ILL}

\newcommand{\stromlo}{\label{stromlo} Research School of Astronomy and Astrophysics, Australian National University, Mt Stromlo Observatory, Weston Creek, ACT 2611, Australia}

\newcommand{\UCatolica}{\label{UCatolica} Instituto de Astronom\'ia, Universidad Cat\'olica del Norte, Av. Angamos 0610, Antofagasta, Chile}

\newcommand{\UT}{\label{UT} McDonald Observatory, The University of Texas at Austin, 1 University Station, Austin, TX 78712-0259, USA}

\newcommand{\Vanderbilt}{\label{Vanderbilt} Department of Physics and Astronomy, Vanderbilt University, VU Station 1807, Nashville, TN 37235, USA}

\newcommand{\UNF}{\label{UNF} Department of Physics, University of North Florida, 1 UNF Dr. Jacksonville FL 32224}

\newcommand{\NAOC}{\label{NAOC} Chinese Academy of Sciences South America Center for Astronomy, National Astronomical Observatories, CAS, Beijing 100101, China}

\newcommand{\CASA}{\label{CASA} Center for Astrophysics and Space Astronomy, University of Colorado, 389 UCB, Boulder, CO 80309-0389, USA}

\newcommand{\UNAM}{\label{UNAM} Universidad Nacional Aut\'onoma de M\'exico, Instituto de Astronom\'ia, AP 106, Ensenada 22800, BC, M\'exico}

\newcommand{\UDP}{\label{UDP} Instituto de Estudios Astrof\'isicos, Facultad de Ingenier\'ia y Ciencias, Universidad Diego Portales, Av. Ej\'ercito Libertador 441, Santiago, Chile}

\newcommand{\Steward}{\label{Steward} Steward Observatory, University of Arizona, 933 N. Cherry Ave., Tucson, AZ 85721-0065, USA}  

\newcommand{\APO}{\label{APO} Apache Point Observatory and New Mexico State University, P.O.\ Box 59,
Sunspot, NM 88349-0059, USA}

\newcommand{\UNAMCU}{\label{UNAMCU} Universidad Nacional Aut\'onoma de M\'exico, Instituto de Astronom\'ia, AP 70-264, CDMX 04510, M\'exico}

\newcommand{\UWash}{\label{UWash} Department of Astronomy, University of Washington, Seattle, WA, 98195}

\newcommand{\CC}{\label{CC} Department of Physics, Colorado College, Colorado Springs, CO 80903}

\newcommand{\Utah}{\label{Utah} Department of Physics and Astronomy, University of Utah, 115 S. 1400 E., Salt Lake City, UT 84112, USA}

\newcommand{\UConcepcion}{\label{UConcepcion} Departamento de Astronom\'ia, Universidad de Concepci\'on, Casilla 160-C, Concepci\'on, Chile}

\newcommand{\FCLA}{\label{FCLA} Franco-Chilean Laboratory for Astronomy, IRL 3386, CNRS and Universidad de Chile, Santiago, Chile}

\newcommand{\Oklahoma}{\label{Oklahoma} Homer L. Dodge Department of Physics and Astronomy, University of Oklahoma, Norman, OK 73019, USA}

\newcommand{\UIUC}{\label{UIUC} Department of Astronomy, University of Illinois, Urbana, IL 61801, USA}

\newcommand{\Harvard}{\label{Harvard} Harvard-Smithsonian Center for Astrophysics, Cambridge, MA 02138, USA}

\newcommand{\caltech}{\label{caltech} Department of Astronomy, California Institute of Technology, Pasadena, CA 91125, USA}

\newcommand{\UOA}{\label{UOA} Department of Physics, University of Arkansas, 226 Physics Building, 825 West Dickson Street, Fayetteville, AR 72701, USA}

\newcommand{\trieste}{\label{trieste} Department of Physics, Astronomy Section, University of Trieste, Via G.B. Tiepolo, 11, I-34143 Trieste, Italy}

\newcommand{\Rad}{\label{Rad}{ Elizabeth S. and Richard M. Cashin Fellow at the Radcliffe Institute for Advanced Studies at Harvard University, 10 Garden Street, Cambridge, MA 02138, USA}}

\newcommand{\UCT}{\label{UCT}{ Department of Astronomy, University of Cape Town, Rondebosch 7701, Cape Town, South Africa}}

\newcommand{\Oxford}{\label{Oxford}{ Sub-department of Astrophysics, Department of Physics, University of Oxford, Keble Road, Oxford OX1 3RH, UK}}


\author{
       K. Kreckel\inst{\ref{HD}} \thanks{\email{kathryn.kreckel@uni-heidelberg.de}}  \and  
       R. J. Rickards Vaught\inst{\ref{STScI}} \and
       O. V. Egorov\inst{\ref{HD}}  \and 
       J. E. M\'endez-Delgado\inst{\ref{UNAMCU}} \and 
       F. Belfiore\inst{\ref{INAF}} \and  
        M. Brazzini\inst{\ref{trieste}} \and
       E. Egorova\inst{\ref{HD}}  \and 
        E. Congiu\inst{\ref{ESOChile}} \and                D.~A.~Dale\inst{\ref{UWyoming}} \and
        S. Dlamini\inst{\ref{UCT}} \and 
    S.~C.~O.~Glover\inst{\ref{ITA}} \and 
        K.~Grasha\inst{\ref{ANU}} \and
    R.~S.~Klessen\inst{\ref{ITA},\ref{IWR},\ref{CfA},\ref{Rad}} \and
        Fu-Heng Liang\inst{\ref{HD}} \and 
        Hsi-An Pan\inst{\ref{TKU}} \and
        Patricia S\'anchez-Bl\'azquez\inst{\ref{UCM}}\and
        Thomas~G.~Williams\inst{\ref{Oxford}} 
}

\institute{\tiny
\HD      \and  
\STScI  \and
\UNAMCU \and
\INAF \and 
\trieste \and
 \ESOChile \and
 \UWyoming \and
\UCT \and
 \ITA \and
 \ANU \and
 \IWR \and
 \CfA \and
 \Rad \and
 \TKU \and
 \UCM \and
 \Oxford 
}

   \date{Received XX; accepted XX}

 
  \abstract
   {Gas-phase abundances provide insights into the baryon cycle, with radial gradients and 2D metallicity distributions tracking how metals build up and redistribute within galaxy disks over cosmic time.}
   {We use a catalog of 22,958 \hii\ regions across 19 nearby spiral galaxies to examine how precisely the radial abundance gradients can be traced using only the \niiauroral\ electron temperature as a proxy for `direct method' metallicities. }
   {Using 534 direct detections of the temperature sensitive \niiauroral\ auroral line, we measure gradients in 15 of the galaxies. Leveraging our large catalog of individual HII regions, we stack in bins of \hii\ region \nii$\lambda$6583 luminosity and radius to recover stacked radial gradients.  }
   {We find good agreement between the metallicity gradients from the stacked spectra,  those gradients from individual regions and those from strong line methods. In addition, particularly in the stacked \te\nii\ measurements, some galaxies show very low ($<$0.05~dex) scatter in metallicities, indicative of a well-mixed ISM. We examine individual high confidence (S/N $>$5) outliers and identify 13 regions across 9 galaxies with anomalously low metallicity, although this is not strongly reflected in the strong line method metallicities. By stacking arm and interarm regions, we find no systematic evidence for offsets in metallicity between these environments, suggesting enrichment within spiral arms is due to very localized processes. }
   {This work demonstrates the potential to systematically exploit the single \niiauroral\ auroral line for detailed gas-phase abundance studies of galaxies. It provides strong validation of previous  results, based on the strong line calibrations, of a well-mixed ISM across typical star-forming spiral galaxies.  }


   \maketitle
%

\section{Introduction}

Chemical abundances are commonly used as a cosmic clock, since heavy elements are only produced within stars and must be returned to the surrounding interstellar medium (ISM) to pollute and enrich the next generation of stars \citep{Maiolino2019}. In the gas-phase, the oxygen abundance (`metallicity') is commonly measured from the rich suite of strong oxygen and hydrogen emission lines produced within photoionized \hii\ regions \citep{Kewley2019}, and detailed study of these metallicities provides a snapshot of the instantaneous metal content in the galaxy at a given position. Radial metallicity gradients are commonly observed in present day galaxies, with enriched gas located towards the galaxy center and the radial decrease well fit by a linear slope \citep{Pilyugin2014, Sanchez2014, Belfiore2017}, sometimes with evidence of inflection points or breaks \citep{Sanchez-Menguiano2016a, Sanchez-Menguiano2018}. While there are different possible physical interpretations for these gradients, they are generally thought to reflect the inside-out growth of galaxies, modified by a complex interplay of outflow-driven removal of metal-rich gas and accretion of metal-poor gas (\citealt{Dalcanton2007, Bresolin2012, Andrews2017}; although see also \citealt{Johnson2024}) along with radial gas flows and other possible internal processes \citep[e.g.][]{Molla2019, Palla2020, Palla2024,Prantzos2023}. In this way, deviations from simple linear trends provide insights into some of the key stellar feedback and cosmological processes that are thought to regulate galaxy evolution. 

Within the past decade, a rich variety of integral field unit (IFU) spectroscopic instruments and new surveys have provided a remarkable view into the 2D gas-phase metallicity distribution as mapped across thousands of individual galaxies (e.g. CALIFA, \citealt{Sanchez2012}; SAMI, \citealt{Bryant2015}; MaNGA, \citealt{Bundy2015}). However at kpc-scales, most of these studies blend \hii\ regions with each other and with surrounding diffuse ionized gas, introducing biases into the inferred metallicities and metallicity gradients \citep{Poetrodjojo2019}.  For smaller samples of nearby (D$<$100~Mpc) galaxies, metallicities can be obtained with much higher precision at the locations of individual \hii\ regions \citep{Rosales-Ortega2011, Blanc2013}, with recent surveys providing measurements for thousands of \hii\ regions per galaxy \citep[e.g.][]{Kreckel2019, Rousseau-Nepton2019, Grasha2022, Groves2023}. With such a high density of metallicity measurements, it has become possible to search for higher-order azimuthal variations in the metal distribution, and connect in detail the sites of metal formation with mixing and diffusion processes, with a particular focus on variations between galactic environments \citep{Ho2017, Kreckel2020, Sanchez-Menguiano2020, Williams2022, Bresolin2025}.  

There are many methods available to infer the metallicity of an HII region from its emission lines \citep{Maiolino2019, Kewley2019}, each of which comes with advantages and disadvantages. Obtaining large statistical samples of metallicity measurements generally requires the use of \emph{strong-line} methods \citep[for a review of various methods see][]{Peimbert2017}. These rely on some combination of bright emission lines (e.g. \ha, \hb, \oii$\lambda$3727, \oiii$\lambda$5007, \nii$\lambda$6583, \sii$\lambda$6717,6731), and make implicit assumptions about the underlying physical conditions of the nebula such as the relation between metallicity, ionization parameter and temperature \citep{Morisset:25}. 
These prescriptions are calibrated either empirically against sets of nebulae where electron temperatures (\te) and densities are measured (the \te-method, defined below), or against photoionization models for a range of conditions (e.g.\ ionization parameter, ISM pressure), to recover a measurement of 12+log(O/H) for a wide range of \hii\ region properties. This is in contrast to the \emph{\te -method} (often called the `direct method'), where collisionally excited, temperature sensitive auroral lines are used in combination with lower energy transitional lines to measure \te\ (e.g.\ \oiii$\lambda$4363 and \oiii$\lambda$5007, \niiauroral\ and \nii$\lambda$6583, \siii$\lambda$6312 and \sii$\lambda$6716,6731). By measuring \te\ it is possible to constrain the expected line emissivities, and recover different ionic abundances (e.g. O$^{+}$/H, O$^{++}$/H). 

The internal structure of \hii\ regions, typically unresolved in external galaxies, arises as radiation from young massive stars ionizes the surrounding gas, creating distinct ionization zones, arranged roughly by their ionization energy.
This layered structure reflects the energy distribution of the radiation field. By measuring lines for multiple ions, it is then possible to obtain abundance measurements for individual ionization zones within each nebula, and infer the total oxygen abundance \citep{Berg2020}. However, these auroral lines are faint, only a few percent of the flux observed in the strong-lines, and hence are typically only observed in the brightest \hii\ regions, requiring dedicated deep observations to recover measurements for samples of $\sim$10s of \hii\ regions per galaxy \citep[e.g.][]{Berg2020}. 
It is worth noting that even these auroral collisionally excited line (CEL) \te-based metallicities are not perfectly understood, and show systematic offsets from the metallicity inferred using even fainter recombination lines (RL) \citep{Esteban:2014, Toribio:2016, Skillman:2020, MendezDelgado:2022a,Chen:2024}. The 0.1--0.3~dex offset between these two physically-grounded methods is known as the abundance discrepancy factor (ADF), which is often attributed to unresolved temperature fluctuations within the nebula \citep{Peimbert1967, jemd_nature}, and reflects the challenges in obtaining a precise understanding of the detailed physics regulating line emission in \hii\ regions. 

Recently, the PHANGS-MUSE survey obtained $\sim$100~pc scale optical IFU coverage of 19 star-forming spiral galaxies \citep{Emsellem2022}, isolating individual star-forming complexes and providing a homogeneous spectroscopic catalog of $\sim$20,000 \hii\ regions \citep{Groves2023}. This includes coverage of the \niiauroral, \sii$\lambda$6312, and \oii$\lambda\lambda$7320,7330 auroral lines.   \cite{Brazzini2024} carried out a careful, tailored fitting of these faint lines, and found 95 \hii\ regions with all three lines detected. \cite{RRV2024} carried out follow-up observations of seven galaxies to provide coverage of the \oiii$\lambda$4363 auroral line, but given the high metallicity of the sample, was only able to obtain detections in 26 (6\%) of the \hii\ regions observed. These intensive efforts to fully map the ionization zones within large samples of nebulae demonstrate the  challenges of building large databases of \hii\ regions where detailed \te-based abundance measurements are possible. 

A simplified approach is possible if one ion could be identified that has a temperature representative of the larger ionized volume. \cite{jemd_nature} used the DESIRED catalog \citep{jemd_desired} to explore temperature relations across a sample of \hii\ regions with exceptionally deep spectra, and found that \te\nii\ alone provides a remarkably accurate constraint on the metallicity across a wide range of physical conditions.  
This is extremely convenient within the context of the PHANGS-MUSE results. 
\cite{Brazzini2024} found that within the $\sim$20,000 PHANGS-MUSE \hii\ regions, \niiauroral\ was the most commonly detected auroral line, with measurements obtained for 969 (about 4\%) of the \hii\ regions. \sii$\lambda$6312 was the least commonly detected, found in $<$1\% of \hii\ regions, and although $\sim$3\% have \oii$\lambda\lambda$7320,7330, the reliability of these detections is questionable as they are heavily blended with bright sky lines in this wavelength range. 

In this work, we focus on exploiting detections of the \niiauroral\ line, as it represents the strongest and best detected auroral line in our sample, to recover direct metallicities of HII regions and investigate radial trends across the disk of our galaxies. We also explore approaches that stack multiple \hii\ region spectra to improve the detection rate of this line and expand our analysis.  Additionally, \nii\,$\lambda$5755/$\lambda$6584 is less susceptible to errors from reddening corrections and telluric emission or absorption, compared to other temperature diagnostics such as \siii\,$\lambda$6312/$\lambda$9069 and \oii\,$\lambda$7330/$\lambda$3728.

In Section \ref{sec:data} we describe the optical spectroscopic data used in this study, the \hii\ region catalog, and our stacking approach. In Section \ref{sec:results} we present our \te\nii\ measurements, from individual and stacked \hii\ regions, present inferred radial metallicity gradients, and explore outliers and environmental variations. In Section \ref{sec:discussion} we validate our choice of strong line prescription for these systems,  consider the utility of \te\nii\ as a direct method abundance diagnostic, and discuss the lack of strong environmental variations. We conclude in Section \ref{sec:conclusion}. 


\section{Data}
\label{sec:data}

\subsection{PHANGS-MUSE spectroscopy}

\begin{table*}
\caption{General properties of the PHANGS-MUSE galaxies.  }
\label{tab:sample}
\centering
\begin{tabular}{lrrrrrr}
\hline \hline

Name & Distance$^{a}$ & $v_\mathrm{sys}^{b}$ & $PA^{c}$ & $i^{c}$ &  $r_{\rm eff}^{d}$ & AO? \\
 & Mpc & km s$^{-1}$ & deg & deg & arcmin &  \\
\hline \hline
IC5332 & 9.0 & 699 & 74.4 & 26.9 &  1.4 & N \\
NGC0628 & 9.8 & 651 & 20.7 & 8.9 &  1.4 & N \\
NGC1087 & 15.9 & 1502 & 359.1 & 42.9 &  0.7 & N \\
NGC1300 & 19.0 & 1545 & 278.0 & 31.8 &  1.2 & Y \\
NGC1365 & 19.6 & 1613 & 201.1 & 55.4 &  3.3$^{e}$ & N \\
NGC1385 & 17.2 & 1477 & 181.3 & 44.0 &  0.7 & Y \\
NGC1433 & 18.6 & 1057 & 199.7 & 28.6 &  0.8 &  Y \\
NGC1512 & 18.8 & 871 & 261.9 & 42.5 &  0.9 & N \\
NGC1566 & 17.7 & 1483 & 214.7 & 29.5 &  0.6 & Y \\
NGC1672 & 19.4 & 1318 & 134.3 & 42.6 &  0.6 & N \\
NGC2835 & 12.2 & 867 & 1.0 & 41.3 &  0.9 & N \\
NGC3351 & 10.0 & 775 & 193.2 & 45.1 &  1.0 & N \\
NGC3627 & 11.3 & 715 & 173.1 & 57.3 &  1.1 & N \\
NGC4254 & 13.1 & 2388 & 68.1 & 34.4 &  0.6 & Y \\
NGC4303 & 17.0 & 1560 & 312.4 & 23.5 &  0.7 & Y \\
NGC4321 & 15.2 & 1572 & 156.2 & 38.5 &  1.2 & Y \\
NGC4535 & 15.8 & 1954 & 179.7 & 44.7 &  1.4 & Y \\
NGC5068 & 5.2 & 667 & 342.4 & 35.7 &  1.3 & N \\
NGC7496 & 18.7 & 1639 & 193.7 & 35.9 &  0.7 & Y \\
\hline
	\multicolumn{7}{p{.5\textwidth}}{ 
		$^{a}${From the compilation of \citet{Anand2021}.}
		$^{b}${From LEDA \citep{Makarov2014}.}
		$^{c}${From \cite{Lang2020}, based on \mbox{CO(2--1)} kinematics.}
            $^{d}${From \cite{Leroy2021}, based on the stellar mass distribution.}
		$^{e}${Due to AGN bias, derived from the  scale length (l$_*$) as r$_{\rm eff}$ = 1.41 l$_*$ following Equation 5 in \citet{Leroy2021}}.
			}
\end{tabular}
\end{table*}

Our sample of 19 star-forming nearby (D$<$20~Mpc) galaxies observed with MUSE \citep{Bacon2010} is listed in Table \ref{tab:sample}. A full description of the observations and data reduction is provided in \cite{Emsellem2022}, and key characteristics of the data set are summarized below.

The MUSE instrument provides optical integral field unit (IFU) spectroscopic imaging across a wide $\sim$1\arcmin\ field of view. The resulting spectra cover $\sim$4800-9300\AA\ at a spectral resolution of R$\sim$2000 \citep{Bacon2017}. In our sample, typical seeing of $\sim$0.7\arcsec\ corresponds to physical scales of 50--100~pc, sufficient to isolate individual \hii\ regions from surrounding nebulae and diffuse ionized gas. As our target galaxies are nearby and extended on the sky, they were typically mosaicked with between 3--15 MUSE pointings to achieve extended coverage of the star-forming disk. 

While all observations were taken using the Wide-Field Mode, eight of the galaxies were observed using Ground Layer Adaptive Optics \citep{Arsenault2008,Strobele2012} to improve the seeing, and increase the spatial resolution. As this employs sodium lasers, a notch filter removes emission from 5820 – 5970 \AA. For galaxies with systemic velocities of more than $\sim$2000~km~s$^{-1}$, depending on the disk rotation this can redshift \niiauroral\ line emission close enough to the notch that accurate recovery of the line flux cannot be guaranteed. This principally affects NGC~4254, and explains the small number of detections in that galaxy.

\subsection{Nebular catalog and \hii\ region selection}
\label{sec:hiiregions}
From the H$\alpha$ emission in these 19 galaxies, a catalog of $\sim$20,000 \hii\ regions were identified and characterized, as detailed in \cite{Santoro2022} and \cite{Groves2023}. Integrated spectra are constructed for each nebula that combine data from all MUSE spaxels associated with the region. These integrated spectra were fit using the penalized PiXelFitting (pPXF)  package \citep{Cappellari2004, Cappellari2017} with a combination of single stellar population (SSP) models (E-MILES, \citealt{Vazdekis2016}) and Gaussian components for strong emission lines (e.g. \hb, \oiii, \ha, \nii, \sii). This results in stellar population models, as well as line flux and line kinematic measurements for each nebula in the catalog. All line fluxes are dereddened by using the \ha/\hb\ line ratio to infer $E(B-V)$, assuming an intrinsic Balmer ratio of \ha/\hb = 2.86, R$_V$ = 3.1, and an \cite{ODonnell1994} extinction curve. Extinction corrected \ha\ luminosities, L(\ha), are calculated assuming the distances listed in Table \ref{tab:sample}. 

Using diagnostic line ratios (e.g. BPT; \citealt{Baldwin1981}), we classify a subset of the nebulae as `\hii\ regions' when we can robustly attribute the line emission to photoionization via the following line ratio criteria. Following \cite{Groves2023}, we require regions to fall below the \cite{Kauffmann2003} diagnostic curve in the \oiii/\hb\ versus \nii/\ha\ BPT diagrams, and below the \cite{Kewley2006} diagnostic curve in the \oiii/\hb\ versus \sii/\ha\ BPT diagrams, with S/N $>$ 5 in all lines.  Where \oi\ is detected with S/N $>$5, we further require that regions fall below the \cite{Kewley2006} diagnostic curve in the \oi/\hb\ versus \sii/\ha\ BPT diagram. We exclude regions that are not fully contained within the field of view and regions overlapping with foreground stars. Unlike in \cite{Groves2023},  to avoid the AO notch filter masking 5820 – 5970 \AA we exclude any regions with measured H$\alpha$ velocities greater than 2345~km~s$^{-1}$. 
Finally, we exclude regions flagged as residing in the galaxy center environment \citep{Querejeta2021}, as they often have unphysically large physical sizes, due to difficulties in deblending clustered star formation within nuclear rings. Taking these criteria into account, we are left with a sample of 20,809 \hii\ regions, with between 367 and 2196 \hii\ regions identified per galaxy. 

The gas-phase oxygen abundance, 12+log(O/H), has been characterized for each of these regions using the \cite{Pilyugin2016} S calibration (Scal), an empirical metallicity prescription that has been shown to provide accurate metallicities for individual \hii\ regions with high precision compared to other prescriptions available in the literature \citep{Ho2019_machine, Metha2021}. Radial gradients for each galaxy were calculated in \cite{Groves2023}, and are in agreement within the reported uncertainties  despite the slightly different \hii\ region selection criteria applied in this work. Subtracting the radial gradient, we can then calculate the offset from the radial gradient, \doh, for each \hii\ region. All galaxies are well characterized by these simple linear radial gradients, which we characterize by calculating \sigmaoh, the standard deviation of \doh\ across the \hii\ regions in the galaxy, finding typical values of 0.03 -- 0.06 dex. This is larger than the typical statistical uncertainties of 0.01 dex that are determined by propagating the errors associated with the line flux measurements. 

As the \niiauroral\ line is typically $\sim$0.5\% of the strength of the \nii$\lambda$6583 line, accurate recovery of the line flux is much more sensitive to the specifics of the stellar population continuum subtraction and the Gaussian line fit.  In this work, we use the auroral line fits from \cite{Brazzini2024}, which were tailored to robustly recover these faint lines. This was done by refitting the stellar populations and including an eighth order multiplicative polynomial to improve the baseline fit, and focusing the Gaussian line fit on a narrow wavelength range around the line of interest while imposing kinematic constraints based on the strong line detections.  This results in 770 of our \hii\ regions with reported \niiauroral\ detections at a S/N$>$3. 

Due to the faintness of the auroral lines, particularly in comparison to the strength of the underlying continuum, a poorly characterized source of uncertainty in our error estimation is due to the difficulty in correctly modeling the underlying stellar population spectrum. This is known to be particularly challenging for the youngest stellar populations (see \citealt{Emsellem2022} and \citealt{Pessa2023}).  Upon inspection, it is apparent that even some of the high-confidence (S/N $>$ 10) line detections are not clearly distinguished in the spectrum containing the stellar continuum emission, even though the stellar spectral templates are expected to be fairly smooth at $\sim$5700~\AA. As we are interested in investigating outliers in the \te\ distribution, we mitigate the poorly parameterized uncertainties associated with the stellar templates by revising our noise estimate to reflect the standard deviation of the original observed spectrum (without stellar continuum subtraction) in two $\pm$50~\AA\ windows neighboring the \niiauroral\ line. This ensures high confidence in our robust (S/N $>$ 5) detections, and reasonable confidence in even the S/N = 3 detections. With these revised errors, our total sample of \hii\ regions with individually detected \niiauroral\ line emission is reduced to  534 \hii\ regions with  S/N $>$ 3, and 213 with S/N $>$ 5. 

\subsection{\hii\ region stacking}
\label{sec:stacking}
The $<$3\% of \hii\ regions where we directly detect \niiauroral\ naturally correspond to the more luminous \hii\ regions in our sample, with $\sim$80\% showing L(\ha) $>$ 10$^{38.5}$ erg~s$^{-1}$, although these still only correspond to $\sim$25\% of the  L(\ha) $>$ 10$^{38.5}$ erg~s$^{-1}$  \hii\ regions. Given the large sample of non-detections with uniform and high quality spectroscopy, this data set is well suited for stacking multiple \hii\ regions together in order to calculate \te\nii\ based on auroral lines detected in the \hii\ region stacks. This allows us to improve the S/N on fainter regions and also to provide more representative averaged measurements for the more luminous regions. 

As we would like to stack regions of similar \te\nii, and hence similar ratios of \niiauroral/\nii$\lambda$6583, we consider bins containing \hii\ regions with matched extinction corrected \nii$\lambda$6583 luminosity, L(\nii). 
In addition, as we know our galaxies to exhibit radial metallicity gradients \citep{Groves2023}, and hence radial \te\ gradients \citep{Kreckel2022}, it is natural to also consider radial bins. We find that steps of $\Delta$\logten(L(\nii)) = 0.5 dex and $\Delta$r = 0.25~r$_{\rm eff}$ generally provide 10 to 100 \hii\ regions across 557 bins.  The specific choice of bin size does not significantly impact our results, and we find these bins provide a good balance of statistics and radial coverage within the galaxies. This includes the \hii\ regions where \niiauroral\ is individually detected to provide representative averaged measurements across the full luminosity range.  

For each spectrum that we consider in the stack, we first subtract the SSP fit, and apply an extinction correction using the same procedure outlined in Section \ref{sec:hiiregions} and adopting the $E(B-V)$ value inferred from the Balmer decrement.  We calculate the rest-frame wavelength by assuming the velocity measured in the \ha\ line (which has higher S/N than the \nii$\lambda$6583 line and shows no significant kinematic offsets), and interpolate the spectrum to a fixed wavelength grid. To account for any residual offset in the continuum level, we subtract the median as calculated between 5700~\AA\ and 5750~\AA.  We consider only bins with at least 5 \hii\ regions, and calculate the mean across all spectra at each wavelength to construct our stacked spectrum. Finally, we  use the \texttt{astropy} \texttt{LevMarLSQFitter} package to fit the spectrum between 5700~\AA\ and 5770~\AA\ with a single Gaussian centered at $\lambda$ = 5754.59\AA. Due to the faintness of the line, we cannot directly constrain the line width as a free parameter. Instead, we measure the 1$\sigma$ line width of the brighter \nii$\lambda$6583 line in the stacked spectrum ($\sim$1.35$\pm$0.05\AA), and adopt this as a fixed line width for the auroral line measurement. Our lines are marginally resolved, and this approach results in $\sim$5-10\% higher line fluxes than would be inferred by adopting the instrumental resolution (1.14 \AA\;   \citealt{Bacon2017}) . Thus, with the mean and $\sigma$ fixed, only the amplitude is fit to provide the measured line flux. The error in the stacked spectrum is calculated by standard error propagation, and included as weights in the Gaussian fit.  
We note that since the input spectra are already dereddened, our resulting auroral line flux measurements are also already extinction corrected. As our stacks are already constructed at fixed values in L(\nii$\lambda$6583), we do not need to measure this directly in the stacked spectrum but instead adopt the mean value as measured in the individual \hii\ regions. This agrees within the uncertainties with what can be measured from the stacked spectrum, and avoids problems arising from deblending \nii$\lambda$6583 from \ha\ and residual stellar absorption features. 
Examples of the stacked spectra and corresponding fits are provided in Appendix \ref{app:stacks}. 

Across our 557 bins that contain at least five \hii\ regions, we recover 215 detections with S/N$>$3, which reflect the properties of 3,599 combined \hii\ regions. 
Compared to the individual detections, the median \logten(L(\nii)/{\rm erg/s/cm}$^2$) in our stacked detections is more sensitive by nearly an order of magnitude, from 38.5 to 37.75. 

\subsection{T$_{\rm e}$ based abundances}
\label{sec:te}

Based on our extinction corrected \niiauroral\ and \nii$\lambda$6583 emission lines, detected in both the individual and stacked \hii\ regions, we use the \texttt{pyneb} package \citep{Luridiana2012, Luridiana2015} to compute  \te\nii. As reported by \cite{Brazzini2024}, for the individual \hii\ regions the function \texttt{getTemDen} was used assuming a fixed \nee\ = 50~cm$^{-3}$ when no robust measurement of \nee\ was available, and \texttt{getCrossTemDen} to solve for both otherwise. They note that none of the \hii\ regions returns a value for \nee\ greater than 200~cm$^{-3}$. For the stacked \hii\ regions, we assume the low-density limit applies and use \texttt{getTemDen} with a fixed \nee\ = 50~cm$^{-3}$. 
As described in several studies (see \citealt{jemd_desired}, and references therein), this \te\ diagnostic is virtually insensitive to density in regions with electron densities below 10,000~cm$^{-3}$—a threshold much higher than all derived densities in our sample.

\begin{figure*}
    \centering
    \includegraphics[width=6.5in]{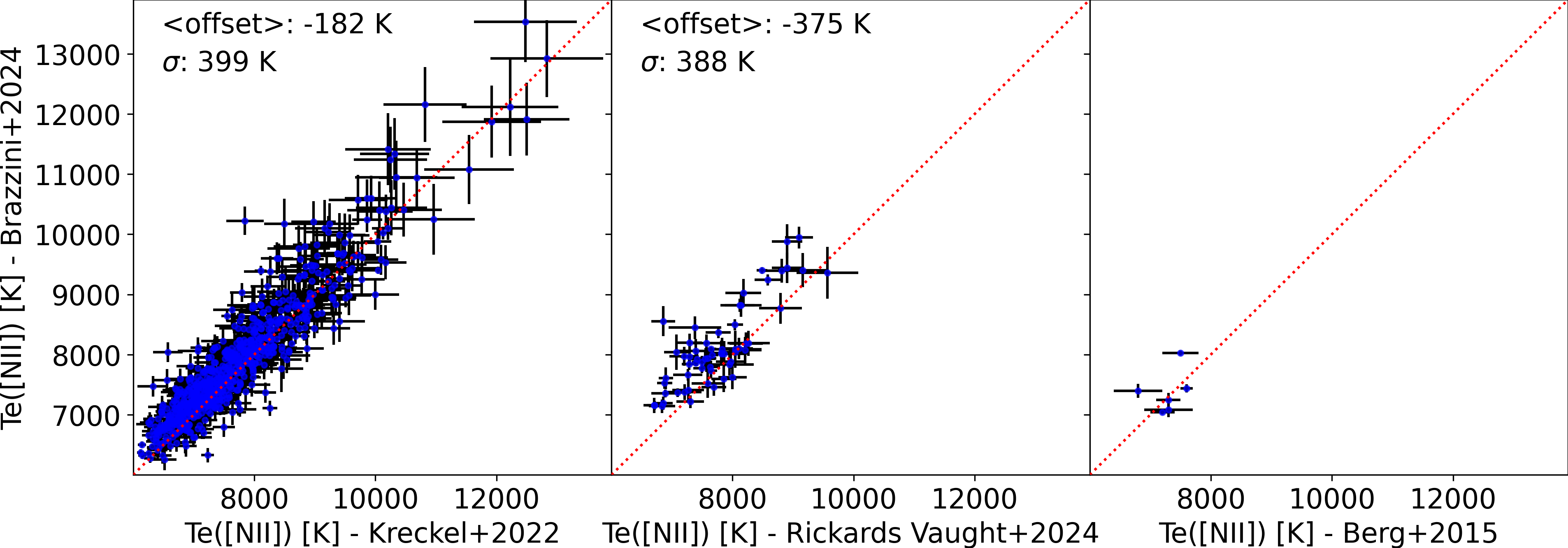}
    \caption{A comparison of \te\nii\ measurements used in this work \citep{Brazzini2024} with three different sets of measurements available in the literature. While auroral line fits in \cite{Kreckel2022} (left) and \cite{RRV2024} (center) are both based on the same underlying MUSE data set, different assumptions have been made in the SSP fitting and region boundaries. \cite{Berg2015} measurements (right) are based on independent long-slit spectroscopy. As there are only five regions in common with \cite{Berg2015}, no offset or scatter is calculated in comparison with this sample.} 
    \label{fig:compare_Te}
\end{figure*}

Figure \ref{fig:compare_Te} presents a comparison of different reported \te\nii\ measurements available in the literature for \hii\ regions in these galaxies, demonstrating the challenges of fitting faint lines (even within the same dataset). \cite{Kreckel2022} used exactly the same input spectra as \cite{Brazzini2024}, but adopted a different set of SSP models. \cite{RRV2024} adopted different region boundaries, resulting in a slightly different integrated MUSE spectrum for each \hii\ region. \cite{Berg2015} used long-slit spectroscopy, only five of which overlap with the MUSE coverage, and again a different set of SSP models. Different works also implemented different Gaussian fitting approaches, with \cite{Berg2015} choosing to calculate line fluxes by integrating over the observed profile linewidth. 
From this comparison, we note that there is good systematic agreement across all measurements, however the overall scatter is on the order of $\sim$400~K, in excess of the typical reported uncertainties of $\sim$100~K. This reflects variations that arise when adjusting region boundaries, stellar template matching and faint line fitting approaches. These uncertainties are not well captured by our reported errors, but should be kept in mind. 

Finally, we convert these measurements of \te\ into 12+log(O/H) using the prescription from \cite{jemd_nature}, 
\begin{equation}
    12+{\rm log(O/H)} = (-1.19\pm0.14) \times T_{\rm e}([{\rm N}\textsc{ii}])/10^4{\rm K} +  (9.68\pm0.15). 
\end{equation}
This relation is based on deep longslit spectra of $\sim$20 Galactic and extragalactic \hii\ regions with temperatures between 8,000-13,000~K, spanning approximately 0.6~dex in metallicty.
This relation is calibrated to abundance measurements made using OII-recombination lines, which has long been known to produce $\sim$0.1-0.2~dex higher measurements of 12+log(O/H) than collisionally excited lines \citep{Wyse1942, Ferland2003, Peimbert2017}, such as [OIII]$\lambda5007$, which are typically used for abundance determinations. This long-standing discrepancy has been the focus of several studies (see \citealt{Stasinska:05, GarciaRojas:07, Henney:10, Ferland:16}, and references therein), which attribute it to various physical phenomena, such as the presence of unresolved temperature fluctuations \citep{jemd_nature}. For these reasons, differences between abundances calculated with the Scal strong-line method (empirically calibrated against collisionally excited lines; see Section \ref{sec:hiiregions}) and this prescription are expected to result in an absolute offset.

\section{Results}
\label{sec:results}

\subsection{T$_{\rm e}$ gradients, in individual \hii\ regions and stacks}
\label{sec:gradients}
\begin{figure*}[ht]
   \includegraphics[width=7in]{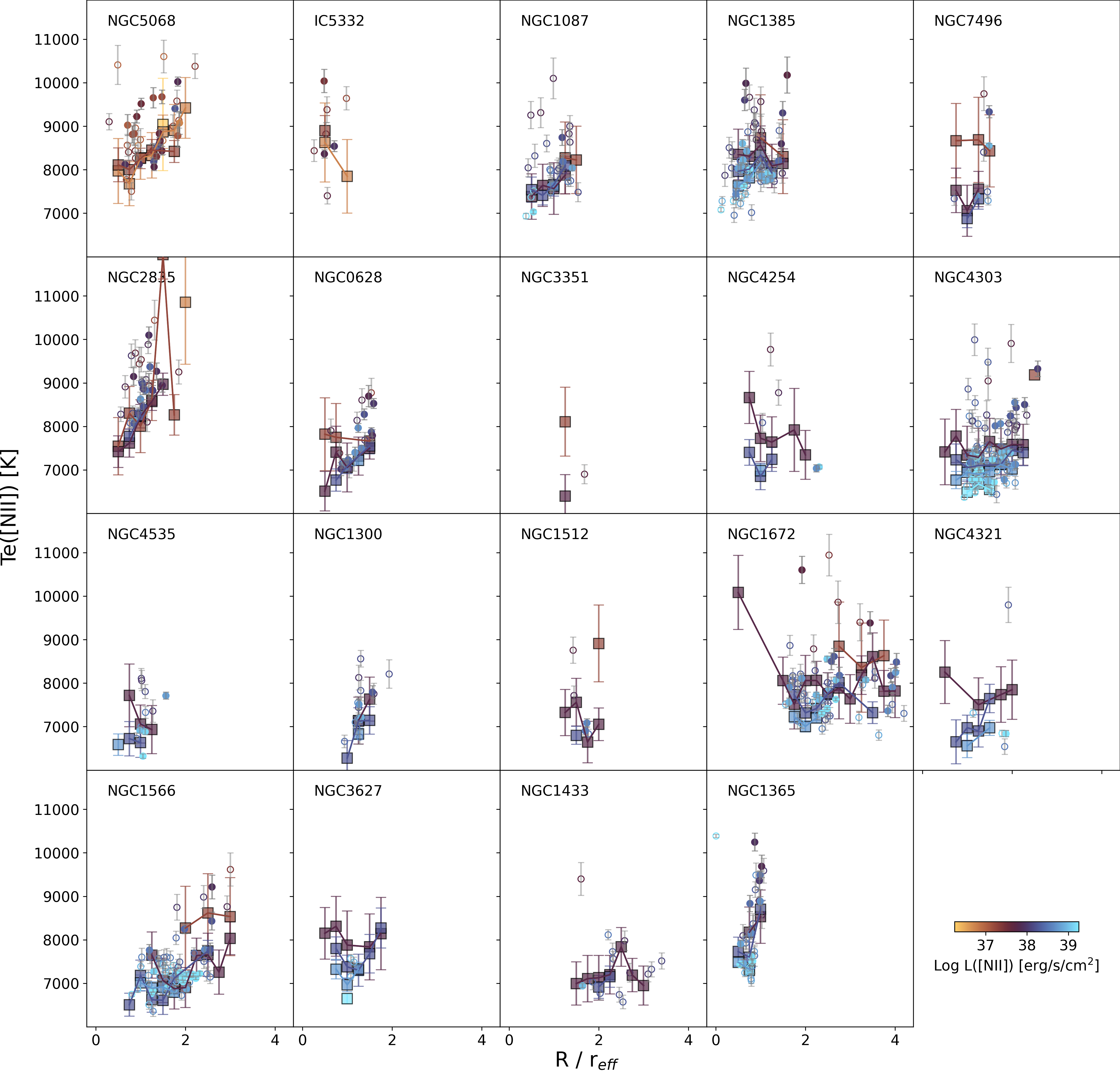}
      \caption{\te\nii\ as a function of radius for all 19 galaxies. We compare individual \hii\ regions (S/N$>$3 open circles, S/N$>$5 filled circles) with measurements from \hii\ region stacks (squares and lines). Points are color coded by their \nii$\lambda$6583 Luminosity (L[\nii]).
      Note that \hii\ regions in NGC~4254 and NGC~4535 do not cover the full disk due to the AO notch filter. Galaxies are ordered from low (top-left) to high (bottom-right) stellar mass, and all galaxies are shown with matched scales, so it is possible to directly compare the absolute values and slopes across the sample. 
              }
         \label{fig:stacks}
\end{figure*}

In Figure \ref{fig:stacks} we show the measured \te\nii\ as a function of radius for each galaxy, comparing the results for individual \hii\ regions with the results obtained from the \hii\ region stacks. Galaxies have been sorted by stellar mass, from low (top) to high (bottom), and all galaxies are shown with matched scales, so it is possible to directly compare the absolute values and slopes across the sample. Higher metallicities are associated with lower temperatures, as metal lines are responsible for cooling, and the overall sample follows the expected mass-metallicity relation \citep{Tremonti2004} with lower mass systems showing systematically higher \te. While many of the radial gradients are quite flat (as was also seen in \citealt{Groves2023}), when trends are present they are typically positive (e.g. NGC~5068, NGC~628, NGC~1566), which corresponds to  expected negative radial metallicity trends.

\begin{figure*}[ht]
   \includegraphics[width=7in]{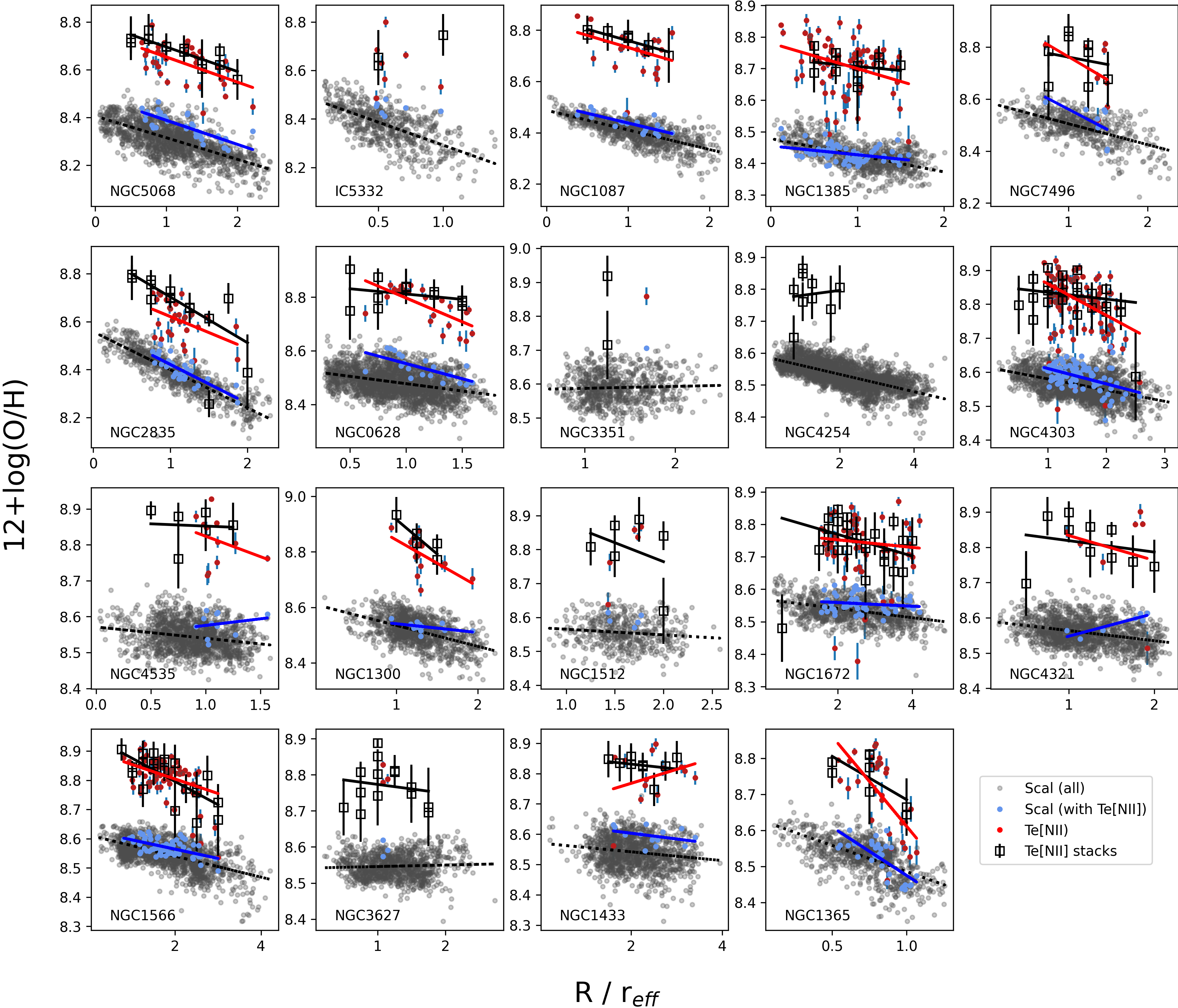}
      \caption{A comparison of metallicity gradients. The Scal values (grey) are compared with individual \te\nii\ metallicities (red) and \te\nii\ stacked metallicities (black). For context, the subsample of individual regions with \te\nii\ detections are highlighted within the Scal measurements in light blue. Linear radial gradients are fit when there are at least 5 measurements that cover at least 0.5~r$_{\rm eff}$. Note that \hii\ regions in NGC~4254 and NGC~4535 do not cover the full disk due to the AO notch filter. Galaxies are ordered from low (top-left) to high (bottom-right) stellar mass, and axis scalings are not  matched  between galaxies.  }
         \label{fig:gradients_scal}
\end{figure*}

All points are colored by L(\nii), and when considering only the individual \hii\ regions there are many galaxies (e.g. NGC~4303, NGC~1566) where a secondary trend in luminosity is suggested. The less luminous regions (in \nii$\lambda$6583) appear to have higher \te\ at fixed radius, in excess of the reported uncertainties. Given the faintness of the \niiauroral\ line, and the fact that higher \te\nii\ derives from a higher intrinsic \niiauroral/\nii$\lambda$6583 line ratio, this is suspiciously suggestive of simple statistical outliers on the high-end tail of the \niiauroral\ line flux distribution that bias us towards overestimating \te\nii. This is supported by the large number of lower 3$<$S/N$<$5 sources with systematically higher \te\ measurements. The stacks confirm this, as while they also show a suggestion of a trend for higher \te\ and lower luminosity, there is significantly less scatter among different luminosity bins at any fixed radial bin (in each galaxy) compared to the individual regions, with temperature differences consistent within the 1$\sigma$ uncertainties.  We revisit the nature of some of the individual high S/N$>$5 outliers in Section \ref{sec:outliers}, but emphasize the good agreement between all stacked \te\ measurements and those from the brightest (and hence highest confidence) \te\ measurements based on individual \hii\ regions. 

Note that in addition to the S/N cuts on the \niiauroral\ line detection, we have also excluded 27 individual regions from our analysis that have reported uncertainties in \te\nii\ of more than 500~K. This is $\sim$3 times higher than the median uncertainty for the sample (146~K), and these 27 regions all have atypically large 
temperatures. These appear to be a combination of spurious spectral features and unusual emission line stars (see Appendix \ref{app:els}).

\subsection{Comparison of metallicity gradients}

\begin{figure*}[ht]
   \includegraphics[width=7in]{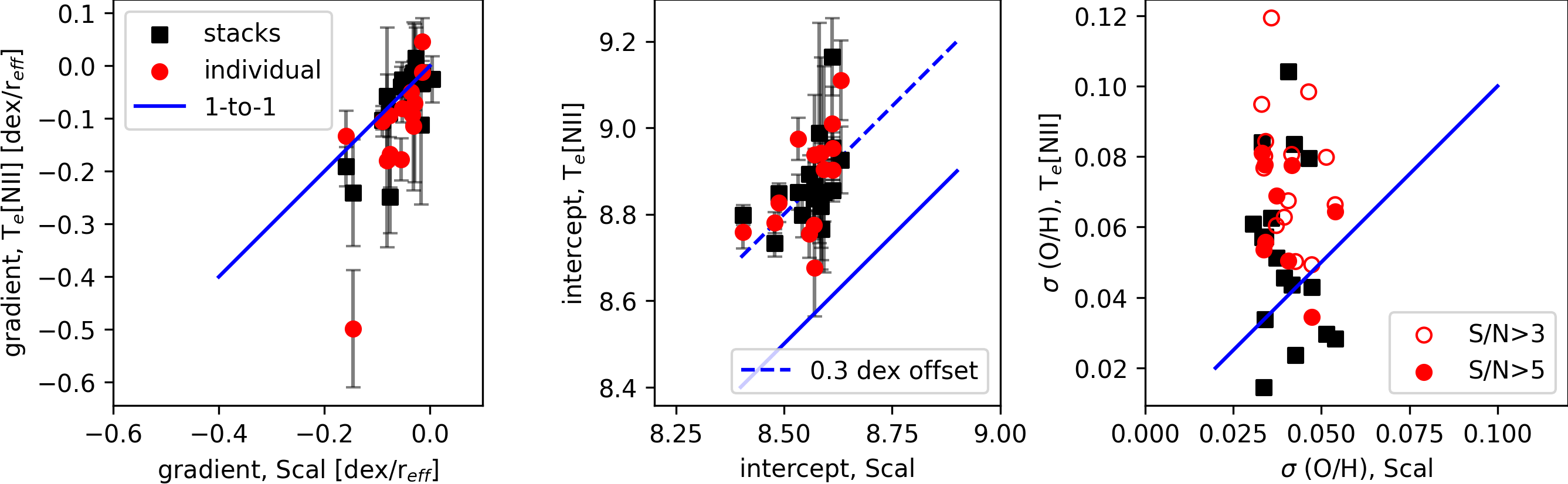}
      \caption{A comparison of the metallicity gradient (left), intercept (center) and scatter (right) between Scal metallicities and the \te\nii\ metallicities in individual \hii\ regions (red) and stacks (black). For reference, the 1-to-1 line is shown, as well as a fixed offset of 0.3~dex for the intercept.  In the right panel, the scatter is measured using individual regions with S/N$>$3 (open) or S/N$>$5 (filled). }
         \label{fig:compare_gradients}
\end{figure*}

Converting our \te\nii\ measurements into a metallicity measurement (see Section \ref{sec:te}), we can directly compare these `direct method' \te\nii\ metallicities with the strong line Scal metallicities derived for the entire \hii\ region population, in both the individual \hii\ regions and the stacked results. 
Figure \ref{fig:gradients_scal} shows this directly for all 19 galaxies, and we note that the axes are rescaled for each galaxy to emphasize the comparison between radial trends in the direct method and strong-line prescriptions. Similar comparisons are made for three additional strong-line prescriptions in Appendix \ref{app:altpresc}. 

We include all \te\nii\ metallicities with S/N$>$3 in our analysis. All radial gradients are fit with outlier removal, using 3$\sigma$ clipping and three iterations. To ensure robust results, gradients are only fit when at least five data points are available, and a range of at least 0.5~r$_{\rm eff}$ is covered by the data. All stacked \hii\ region measurements for a given galaxy are considered together to obtain a \te\nii\ stack radial gradient for that galaxy. 

Qualitatively, we see very good agreements between the slopes in all fits. As explained in Section \ref{sec:te}, the absolute offset between the Scal and \te\nii\ based metallicities is expected, and reflects the abundance discrepancy factor between abundances derived from collisionally excited lines compared to recombination lines. 
The metallicity of the \te\ stacks is in good agreement with what is observed in individual \hii\ regions. Our stacking approach also consistently achieves detections towards the galaxy centers where this low-temperature, high-metallicity regime is frequently inaccessible for individual regions without exceptionally deep spectroscopy.

NGC~4254 has sufficiently high systemic velocity that the \niiauroral\ line detection is severely impacted by the AO filter notch, explaining the poor recovery of \te\nii\ metallicities. Other galaxies with poorly sampled \te\nii\ metallicities (IC~5332, NGC~1512, NGC~3351) have overall smaller \hii\ region populations and limited radial coverage within the field of view, limiting the effectiveness of our statistical stacking approach.

As the individual \hii\ regions with \te\nii\ measurements represent a small fraction of the total \hii\ region population, we isolate and separately fit the Scal metallicities for that exact subset of \hii\ regions (light blue points), to also consider if their Scal gradients accurately reflect the entire population. In general, we see good agreement, particularly in those galaxies with a large number ($>$20) of \hii\ regions with \niiauroral\ detections. We do note that in some galaxies (e.g. NGC~5068, NGC~628, NGC~7496, NGC~1433) the subsamples with \te\nii\ detections are offset towards higher Scal metallicties, but by only a small amount $<$ 0.1~dex. This appears to depend on the calibration adopted (see Appendix \ref{app:altpresc} and figure 13 in \citealt{Brazzini2024}). This could be related to the effect noted in \cite{Kreckel2019}, where more luminous \hii\ regions with higher ionization parameter are correlated with higher offsets from the radial trend (\doh), which correspond to localized pockets of enriched material around vigorously star-forming regions that have not become well mixed over $>$100~pc scales. 

We provide a direct comparison of the metallicity gradient, intercept and scatter about that gradient for the different linear fits in Figure \ref{fig:compare_gradients}. Both the gradient from the \te\nii\ stacks and the individual \hii\ regions follow the 1-to-1 trend compared to the Scal gradient, demonstrating that the Scal metallicities robustly trace the relative changes in metallicity that are recovered from the \te\ direct method metallicity. The offset in the intercept is well explained by the ADF, and corresponds to an absolute underestimate of the Scal metallicities by $\sim$0.3~dex compared to the recombination line calibrated \te\nii\ metallicities. 

One of the surprising characteristics of the Scal metallicity gradients reported by \cite{Kreckel2019} and confirmed in other galaxies \citep{Bresolin2025} is the very small \sigmaoh\ of 0.03-0.06~dex routinely measured, which is even smaller when considering \sigmaoh\ within $<$ kpc scale sub-regions of the galaxies \citep{Kreckel2020}. This is effectively the uncertainty inherent to the prescription, and suggests a high level of homogeneity (perhaps due to efficient mixing) in gas-phase abundances. 
From the \te\nii-based metallicities of the individual \hii\ regions with S/N$>$ 3, we see slightly higher \sigmaoh\ ($\sim$0.08~dex), but still almost exclusively measure variations of less than 0.1~dex. We note that the \te\nii-based metallicities in the stacks show consistently higher levels of homogeneity, with \sigmaoh\ well matched to the Scal derived values. If the variations in O/H are driven partly by luminosity or azimuthal location, then a possible reason for the stacks to show lower scatter is that we are averaging over the driving variable. However, we also find that \sigmaoh\ is systematically lower when considering only the high confidence (S/N$>$5) individual detections. As our observed \sigmaoh\ corresponds to \te\ variations of $\sim$700~K, while the typical errors in \te\ are only $\sim$150~K, this could also arise if errors in \te\nii\ are routinely underestimated, propagating into higher reported values of \sigmaoh. The possibility of systematic error underestimation is supported by our measurement of higher than expected scatter when comparing different \te\nii\ measurements from the literature for the same set of \hii\ regions (Figure \ref{fig:compare_Te}), which show $\sim$400~K scatter arising from variations in the treatment of the underlying SSP and region size when measuring the auroral line fluxes.

\subsection{Metallicity outliers}
\label{sec:outliers}

Given the clear radial gradients in \te\nii\ for many of the galaxies, and large number of individual \hii\ region detections, it is possible to look specifically at outliers in these distributions, to try and understand if these are indicative of significant azimuthal abundance variations. 
We identify a total of 13 \hii\ regions in nine galaxies with 12+log(O/H) values based on the \te\nii\ measurements that are high confidence detections (S/N $>$ 5) and are at least 0.1~dex offset below the radial trend, and two regions at least 0.1~dex offset above the radial trend. Table \ref{tab:outliers} lists the position of all regions, along with the host galaxy name and region ID as given in \cite{Groves2023}, and the \te\nii\ metallicity measured for the individual region as well as the median value across all luminosities of the corresponding radial stack.  

These outlier \hii\ regions are generally $\sim$1~$\sigma$ offset from the radial gradient measured in the individual \te\nii\ metallicities ($\Delta$(O/H)). In comparison to the corresponding stack at matched radius, the low metallicity outliers similarly pronounced, with values 0.2-0.3~dex lower than in the stacks. The high metallicity outliers appear much more consistent with the stacked results ($<$0.1~dex offsets). 

The \hii\ region outliers in metallicity do not show any strong kinematic signatures distinguishing them from the bulk rotational motion of the disk as seen in surrounding ionized gas, which might have been expected if these differences were due to external gas accretion effects. Of the regions showing lower metallicities, three are located on spiral arms, three are located in the `disk' environment (typically where no pronounced spiral structure is seen) and seven are in interarm environments. Of the regions with higher metallicities, one is in the `disk' and the other is in an interarm environment. Although these are very small number statistics, the interarm environment appears slightly over-represented (see also Section \ref{sec:env}), as they contain only $\sim$35\% of all \hii\ regions in these galaxies. Two of the regions with low metallicity in NGC~1385 are located next to each other, with a separation of $\sim$100~pc. 

\begin{figure}
    \centering
    \includegraphics[width=0.95\linewidth]{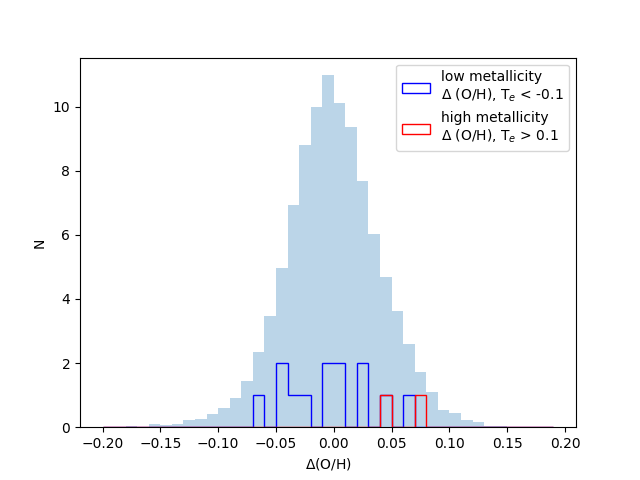}
    \caption{The distribution of $\Delta$(O/H) based on the Scal metallicities for the full \hii\ region sample (filled blue). This is compared with the offsets measured from the strong-line Scal prescription for \te\nii\ outliers to low (blue lines) and high (red lines) metallicities. The full \hii\ region sample distribution has been normalized, while the outlier distribution reflects direct counts. }
    \label{fig:hist_doh}
\end{figure}

As it remains challenging to recover \te\nii\ measurements for large numbers of individual \hii\ regions, it is useful to examine whether these metallicity outliers are also apparent via their strong-line metallicities. Figure \ref{fig:hist_doh} compares \doh\ in our 15 outliers, as measured from the Scal prescription and radial gradients, with the full \hii\ region sample. While the two outliers to high metallicity (in red) are also modest outliers in \doh$_{\rm Scal}$, the distribution for the low metallicity outliers is quite flat. This suggests that, if these individual measurements of high \te\nii\ are real, then it is not possible to easily identify them from strong-line metallicity prescriptions alone. Inconsistencies when attempting to identify metallicity outliers have also been seen when comparing strong-line methods (see Appendix C in \citealt{Kreckel2019}). Given large enough sample sizes, by design a small fraction will appear as 3$\sigma$ outliers in the distribution, however these inconsistencies (particularly for low metallicity outliers) suggest more stringent criteria should perhaps be applied before interpreting these in the context of pristine gas accretion. 

Finally, we note two extreme outliers (excluded from Table \ref{tab:outliers} due to large uncertainties in \te\nii) with reported temperatures that are unphysically large for \hii\ regions (30,000~K and 70,000K), but convincing \niiauroral\ detections. However, these appear to be associated with Wolf-Rayet or other emission-line stars (see Appendix \ref{app:els}), and detailed analysis of these sources is beyond the scope of this work.

\begin{table*}
\caption{Locations and metallicities for \hii\ regions with high and low reported \te\nii\ metallicities.  }
\label{tab:outliers}
\centering
\begin{tabular}{lrrr|rr|rr}
\hline \hline
Galaxy & region ID & RA [hms] & Dec [dms] & \multicolumn{4}{c}{12+log(O/H)} \\
 & & [h:m:s] & [d:m:s] &  \multicolumn{2}{c}{\te} & \multicolumn{2}{c}{Scal} \\
 & & & & individual & $\Delta$(O/H) & individual &  $\Delta$(O/H) \\
\hline
\multicolumn{7}{l}{Low metallicity outliers} \\
\hline
NGC1365  & 188 &  03:33:27.42 & -36:09:12.50  &  8.46  & -0.22 &  8.46 & -0.05 \\
NGC1365  & 352 &  03:33:31.86 & -36:07:07.61  &  8.63  & -0.10 &  8.52 & -0.00 \\
NGC1385  & 46 &  03:37:30.80 & -24:30:52.12  &  8.47  & -0.18 &  8.43 & 0.03 \\
NGC1385  & 333 &  03:37:28.20 & -24:29:38.14  &  8.49  & -0.23 &  8.39 & -0.04 \\
NGC1385  & 717 &  03:37:28.31 & -24:29:38.82  &  8.54  & -0.19 &  8.39 & -0.04 \\
NGC1566  & 12 &  04:20:10.36 & -54:56:23.30  &  8.58  & -0.19 &  8.51 & -0.01 \\
NGC1566  & 266 &  04:20:07.48 & -54:56:37.11  &  8.70  & -0.11 &  8.55 & 0.00 \\
NGC1672  & 75 &  04:45:28.26 & -59:14:11.80  &  8.56  & -0.17 &  8.55 & 0.03 \\
NGC1672  & 1372 &  04:45:34.86 & -59:14:16.40  &  8.42  & -0.33 &  8.52 & -0.02 \\
NGC2835  & 723 &  09:17:52.25 & -22:20:12.14  &  8.48  & -0.12 &  8.42 & 0.05 \\
NGC4303  & 80 &  12:22:00.08 & +04:29:26.57  &  8.57  & -0.14 &  8.53 & 0.00 \\
NGC5068  & 861 &  13:18:51.80 & -21:01:13.72  &  8.55  & -0.10 &  8.38 & 0.07 \\
NGC7496  & 259 &  23:09:46.53 & -43:26:41.77  &  8.57  & -0.11 &  8.41 & -0.06 \\
\hline
\multicolumn{5}{l}{High metallicity outliers} \\
\hline
NGC1433  & 335 &  03:41:56.69 & -47:14:18.17  &  8.85 & 0.10 & 8.62 & 0.08  \\
NGC4535  & 1004 &  12:34:20.68 & +08:10:26.69  &  8.93 & 0.11 & 8.58 & 0.04  \\
\hline
\hline
\end{tabular}
\end{table*}

\subsection{Comparison of spiral arm and interarm environments}

Our stacking approach enables us to determine the bulk properties of \hii\ regions at fixed radius and luminosity within each of our galaxies. Our sample size is large enough that we can take this a step further and also separate the sample by galactic environment. Many recent studies have identified differences in the metallicity of \hii\ regions on spiral arms compared to interarm environments \citep{Kreckel2019, Sanchez-Menguiano2020}, although the trends are not strong and are not well reflected in the bulk distribution of metals across the entire environment \citep{Williams2022}. 

\begin{figure}
    \centering
    \includegraphics[width=1\linewidth]{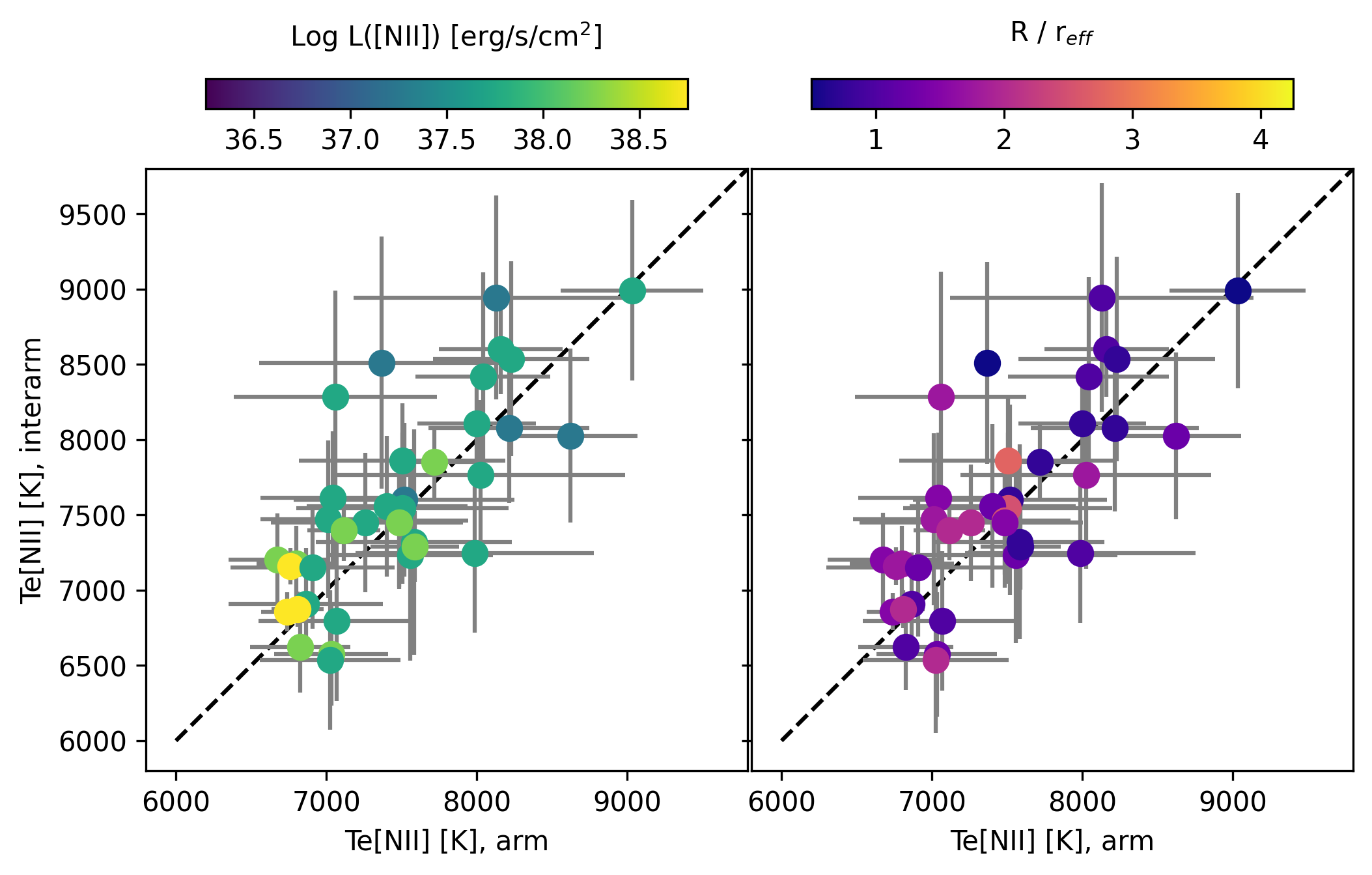}
    \caption{A comparison of \te\nii\ measured in stacked arm and interarm environments. We color-code the points by L(\nii) (left) and radius (right) but see no evidence for trends or bulk offsets between environments. }
    \label{fig:stack-arm-interarm}
\end{figure}

We split our stacks in radius and luminosity into two further bins, reflecting the arm and interarm environments as defined by \cite{Querejeta2021}. Here, we consider only those environments that are outside of the stellar bar (if present), and exclude galaxies where no spiral structure was clearly identified (six galaxies). For stacks where both environments have at least 5 \hii\ regions, we can then directly compare \te\nii\ measurements across both environments for the bulk \hii\ region population. Figure \ref{fig:stack-arm-interarm} shows that both environments agree within $\sim$500~K across the full range of \nii\ luminosities and radii that we cover. We see no evidence of secondary trends or offsets with either parameter. 

\section{Discussion}
\label{sec:discussion}

\subsection{Comparison of Scal strong-line and \te\nii\ direct method metallicities} 

Our comparison of direct method \te\nii\ metallicities with strong-line Scal metallicities has demonstrated a good agreement between the derived radial gradients (Figure \ref{fig:compare_gradients}), particularly when improving the S/N of the faint \niiauroral\ auroral line by stacking in radial bins. This provides confirmation that the trends identified within our order of magnitude larger samples of \hii\ regions based on strong-line methods are an accurate reflection of trends in the metallicity. Note that our comparison is limited to the relatively high metallicities (from half-solar to solar) covered by our sample. 

A comparison with other strong-line prescriptions is provided in Appendix \ref{app:altpresc}. 
Overall, the metallicity gradient slopes for all prescriptions correlate broadly with the \te\nii\ metallicities, with the O3N2 and  N2 prescriptions showing the largest deviations and an overall narrower dynamic range in measured slopes. 
We observe that the N2S2 \citep{Dopita2016} calibration behaves similarly to the Scal prescription, which is unsurprising given the strong correlation between metallicities in these two prescriptions \citep{Groves2023}. As the N2S2 metallicities are typically higher, the absolute metallicity values are also in slightly better agreement with our \te\nii\ metallicities.  

The low $<$0.1~dex scatter in metallicity that is seen within galaxies relative to their radial gradients, even in the individual \te\nii\ metallicity measurements, reflects that a systematic, kpc-scale mixing must act to homogenize the metal distribution within galaxy disks. Most of the scatter in the individually detected \te\nii\ metallicities is associated with low confidence 3 $<$ S/N $<$ 5 \hii\ regions, and could likely be spurious. Increasing the threshold to include only S/N $>$ 5 regions, our statistics become limited but we do see that the scatter drops to 0.1--0.05~dex, similar to what is found in the Scal metallicities. Again, this provides confirmation that the results obtained on the relative metallicity variations using the Scal metallicities is robust. This high level of homogeneity has been reported in other individual galaxies \citep[e.g.][]{Li2013, Croxall2016, Rogers2022}, and with our galaxy sample we are able to demonstrate that this is likely a general characteristic of spiral galaxies.

The absolute offset between the \te\nii\ and Scal metallicities is approximately 0.3~dex across the full range of metallicities probed by our sample. This is due to the ADF, as our prescription for calculating \te\nii\ metallicities is calibrated based on recombination lines instead of collisionally excited lines \citep{jemd_nature}.  A detailed comparison of strong-line metallicities and \te\nii\ measurements for individual regions might reflect differences in unresolved spatial fluctuations in temperature  between \hii\ regions, but is beyond the scope of this work. In the calibration of \te\nii\ as a metallicity indicator, no secondary dependence on unresolved spatial fluctuations in temperature 
was seen \citep{jemd_nature}, supporting the case for using this as a robust single-temperature indicator of gas-phase metallicity.  

One feature of note in the radial metallicity gradients shown in Figure \ref{fig:gradients_scal} is that for some galaxies (e.g. NGC~5068, NGC~628, NGC~1433) the individual \hii\ regions with \te\nii\ measurements (blue points) are not representative of the full \hii\ region sample when considering their Scal metallicities. They are systematically offset towards higher metallicities by almost 0.1~dex. This effect is not present in all galaxies, and does not correlate with absolute metallicity or any other obvious property (e.g. stellar mass, star formation rate, diffuse ionized gas contribution). As the \niiauroral\ line is faint, these are preferentially the more luminous \hii\ regions in each galaxy, and this effect is reminiscent of the correlations reported in \cite{Kreckel2019} for \hii\ regions with higher ionization parameter, higher \ha\ luminosity, and younger stellar clusters to be associated with localized enrichment (higher \doh). Considering only the \te\nii\ metallicities, the smaller number of \hii\ regions and overall larger scatter makes it hard to determine if a similar trend is present. In the \hii\ region stacks, any trends in luminosity are minimized, suggesting that such a trend with luminosity in the Scal metallicities may also be spurious. If so, it would also imply that the radial gradients are even more homogeneous, as \doh\ measured with the Scal prescription reduces to $<$0.03~dex when considering the sub-sample of \hii\ regions where \niiauroral\ is detected. 

\subsection{\te\nii\ as a new standard?}
The use of `direct method' metallicities, where \te\ is directly measured in order to constrain ionic abundances, has long been seen as a robust method of measuring metallicities in \hii\ regions \citep{Peimbert2017}. Given the ionization structure of \hii\ regions, the most precise approach comes from combining multiple temperatures from multiple ions, constructing 2- or 3-zone models that can be used to measure both O$^{+}$ and O$^{++}$ abundances, to compute a total integrated O/H abundance \citep[e.g.][]{Garnett1992, Bresolin2009, Berg2015}. When measurements are not available for specific ions, temperature-temperature relations are commonly used to account for information on the missing ionization zones \citep[e.g.][]{Campbell1988, Berg2020}. However, this detailed approach still relies on assumptions about the nebular conditions (e.g. density; \citealt{jemd_desired}), and by using multiple ionization zones it compounds the line flux uncertainties for multiple faint auroral lines.

The canonical approach comes from measuring \te\oiii\ using the \oiii$\lambda$4363 auroral line. However, this line is particularly faint in high metallicity systems, where it can also be blended with [FeII] lines such as $\lambda 4360$  \citep{Curti2017, Rogers2022}, and recent work has suggested it may be further biased by shocks arising from localized stellar feedback \citep{RRV2024, Khoram2025}. 
Constraining O$^{+}$ requires detection of the \oii$\lambda$7320,7330 auroral lines, which are blanketed by sky lines, and also widely separated from the \oii$\lambda$3727 strong-line (also needed to constrain O$^{+}$) and hence sensitive to errors in the extinction correction. 
In addition, both of these oxygen temperatures rely on emission lines located at blue wavelengths that are not covered by  VLT/MUSE, limiting the statistics on these lines until the advent of new instruments like BlueMUSE \citep{Richard2019}.  
When these oxygen temperatures are not available, a common alternative is to substitute temperatures from different ions that are expected to populate a similar ionization zone in the nebula. For example, \te\siii\ provides an alternative for \te\oiii, although it only partly spans the O$^{++}$ zone and has strong line and auroral lines that are again widely separated in wavelength.

\cite{jemd_nature} demonstrated that \te\nii\ provides a remarkably good representation of \te\ across the entire ionized nebula, and has the added advantage of appearing insensitive to unresolved temperature fluctuations. There is only a modest wavelength separation between the \niiauroral\ auroral line  and \nii$\lambda$6583 strong line, relatively little sky line contamination, and it is very well suited to stacking approaches. 
In this work, we have explored the use of the sole \te\nii\ measurement to infer metallicities across 534 \hii\ regions and 215 stacks, and found good systematic agreement with the results derived from the strong-line Scal prescription. 

We highlight two important caveats to our result. First, we have only tested this approach in the high-metallicity regime (8.2 $<$ 12+log(O/H) $<$ 9.0), and \niiauroral\ remains challenging to detect in low-metallicity systems. However, the range of metallicities over which \te\nii\ provides robust measurements includes virtually all spiral galaxies, highlighting its potential  for investigating gradients and mixing in massive galaxies. 
Second, any detailed dependence of the conversion between \te\nii\ and 12+log(O/H) on the ionization structure of the region (i.e.\ ionization- vs density-bounded regions, Wolf-Rayet winds or extreme shocked interior conditions, blister-like morphology) has not yet been explored. 
In the case of such realistic complications, then the relative contribution of O$^{++}$ to the total abundance could differ from the reference sample used in \cite{jemd_nature}, and a 1-zone approach might fail. 
However, we believe \te\nii\ holds significant promise as a new gold standard worthy of more detailed exploration, as it could serve as a foundation for the development of improved strong-line prescriptions.

\subsection{The challenge of detecting higher-order or environmental metallicity variations}
\label{sec:env}

IFU surveys of nearby galaxies are moving us from radial studies of metallicity, with a small number of \hii\ regions, towards a more comprehensive 2D map of the `metal field'. These higher-order patterns in the metal distribution are important to understand, as they hold information on the balance of feedback, accretion, transport and winds in enriching the ISM \citep{Sharda2024}. 
Separation of trends by galaxy environment \citep{Ho2017, Kreckel2019, Sanchez-Menguiano2020, Bresolin2025} mirror some effects seen in some simulations \citep{Grand2016, Molla2019, Spitoni2019, Spitoni2023,Khoperskov2023, Orr2023}, though sometimes with less pronounced (or absent) differences between spiral arm and interarm metallicities \citep{Grasha2022}. Interpolating metallicity measurements between \hii\ regions is also possible by assuming a smooth 2D underlying metal distribution \citep[e.g.][]{Metha2022}, but so far this also fails to recover any systematic environmental or other higher-order trends beyond a linear radial gradient \citep{Williams2022}.  However, concerns about biases to metallicity diagnostics remain, particularly due to the impact of diffuse ionized gas blended with fainter regions \citep{Poetrodjojo2019, Perez-Montero2023, Lugo-Aranda2024, Gonzalez-Diaz2024}.

\cite{Ho+2019} demonstrated that metal enrichment along the spiral arm ridge of NGC~1672 is well recovered via azimuthal \te\nii\ variations, however that one galaxy remains an outlier (possibly due to its strong bar or well separated spiral arms) in demonstrating such a pronounced environmental trend. As seen in Section \ref{sec:outliers}, the extreme outliers in \te\nii\ show no strong environmental preference, with only a hint that lower metallicity regions are found more often in the interarm. By stacking all regions at matched luminosity and radius by environment, in Figure \ref{fig:stack-arm-interarm} we also see no pronounced difference due to environment alone.  As the variation in NGC~1672 is also not recovered by this analysis, we believe this reflects how localized ($<$300~pc) these metallicity variations are, which has also been suggested in studies using strong-line metallicities \citep{Kreckel2020}, such that they are not easily captured by $\sim$kpc scale environmental masks. 

Note that our analysis excludes entirely any consideration of the galaxy centers. These environments have shown evidence for unusual stellar abundances \citep[e.g.][]{Sextl2024}, but gas-phase studies remain challenging due to the high level of crowding of \hii\ regions, and the potential contribution to line ionization by low-mass stars (given the high stellar density), shocks (from bar-driven gas flows), or AGN excitation.

\section{Conclusions}
\label{sec:conclusion}

We have used PHANGS-MUSE spectroscopy to examine \te\nii\ measurements in 534 individual \hii\ regions and 215 stacked regions across 19 nearby galaxies. Using the prescription from \cite{jemd_nature}, we convert these to measurements of the metallicity. 
After accounting for the instrumental masking of specific wavelength ranges due to the AO system, and requiring radial coverage of more than 0.5~r$_{\rm eff}$, we are able to constrain the metallicity gradients in 15 of our target galaxies. 
We find very good agreement between the radial gradients when compared to strong-line Scal metallicities, but a large 0.3~dex intercept offset due to the abundance discrepancy factor. Our stacking approach considers bins in fixed radius and \nii$\lambda$6583 luminosity, and we can demonstrate that much of the scatter in \te\nii\ metallicities is due to low S/N$\sim$3 detections of \niiauroral\ in individual regions that introduce biases towards high \te\nii\ values. At high S/N $>$ 5, we find very homogeneous gradients in our \te\nii\ metallicities across each galaxy (\sigmaoh $\sim$0.1--0.05~dex). From our stacked data, we see consistent results across different luminosity bins, and very little deviation from a linear radial trend, suggesting efficient azimuthal mixing. 

Using the individual \hii\ region \te\nii\ metallicities, we search for outliers at the $\sim$0.1~dex level, and identify 13 regions with anomalously low metallicity and two regions with anomalously high metallicity. These deviations are not well reflected in the corresponding strong-line metallicities, and show no clear correlations with disk kinematics (which might indicate connections with gas flows or external accretion) or environment (e.g. arm vs interarm). By stacking \hii\ regions separately by arm and interam environments, we also see no large kpc-scale systematic differences in the abundance, suggesting that previously identified enrichment along spiral arm ridges must be due to localized processes.

This work tests a temperature-based approach to measuring metallicities that relies on only a single auroral line detection. By demonstrating good systematic agreement with the well-used Scal strong-line prescription \citep{Pilyugin2016}, this serves to cross-validate the two methods for the metallicitiy regime probed by our galaxies. We find the use of \te\nii\ demonstrates significant promise for future studies of precise abundance determinations within nearby galaxies.

\begin{acknowledgements}
We thank the referee for helpful comments that improved this work.

KK, EE and FHL gratefully acknowledge funding from the Deutsche Forschungsgemeinschaft (DFG, German Research Foundation) in the form of an Emmy Noether Research Group (grant number KR4598/2-1, PI Kreckel) and the European Research Council’s starting grant ERC StG-101077573 (“ISM-METALS"). 

SCOG and RSK acknowledge financial support from the European Research Council via the ERC Synergy Grant ``ECOGAL'' (project ID 855130) and from the German Excellence Strategy via the Heidelberg Cluster of Excellence (EXC 2181 - 390900948) ``STRUCTURES''.

OE acknowledges funding from the Deutsche Forschungsgemeinschaft (DFG, German Research Foundation) -- project-ID 541068876.

Table \ref{tab:sample} includes distances that were compiled by \citet{Anand2021} from \citet{Freedman2001,Nugent2006,Jacobs2009,Kourkchi2017,Shaya2017,Kourkchi2020} and \citet{scheuermann2022}. 

Based on observations collected at the European Southern Observatory under ESO programmes 094.C-0623 (PI: Kreckel), 095.C-0473,  098.C-0484 (PI: Blanc), 1100.B-0651 (PHANGS-MUSE; PI: Schinnerer), as well as 094.B-0321 (MAGNUM; PI: Marconi), 099.B-0242, 0100.B-0116, 098.B-0551 (MAD; PI: Carollo) and 097.B-0640 (TIMER; PI: Gadotti). 

This research has made use of the NASA/IPAC Extragalactic Database (NED) which is operated by the Jet Propulsion Laboratory, California Institute of Technology, under contract with NASA. It also made use of a number of python packages, namely the main \textsc{astropy} package \citep{Astropy+2013, Astropy+2018, Astropy+2022}, \textsc{numpy} \citep{Harris+2020} and \textsc{matplotlib} \citep{Hunter+2007}.
\end{acknowledgements}

\bibliographystyle{aa_url}

\begin{appendix} 

\section{Example stacked spectra}
\label{app:stacks}

\begin{figure*}
\centering
\includegraphics[width=7in]{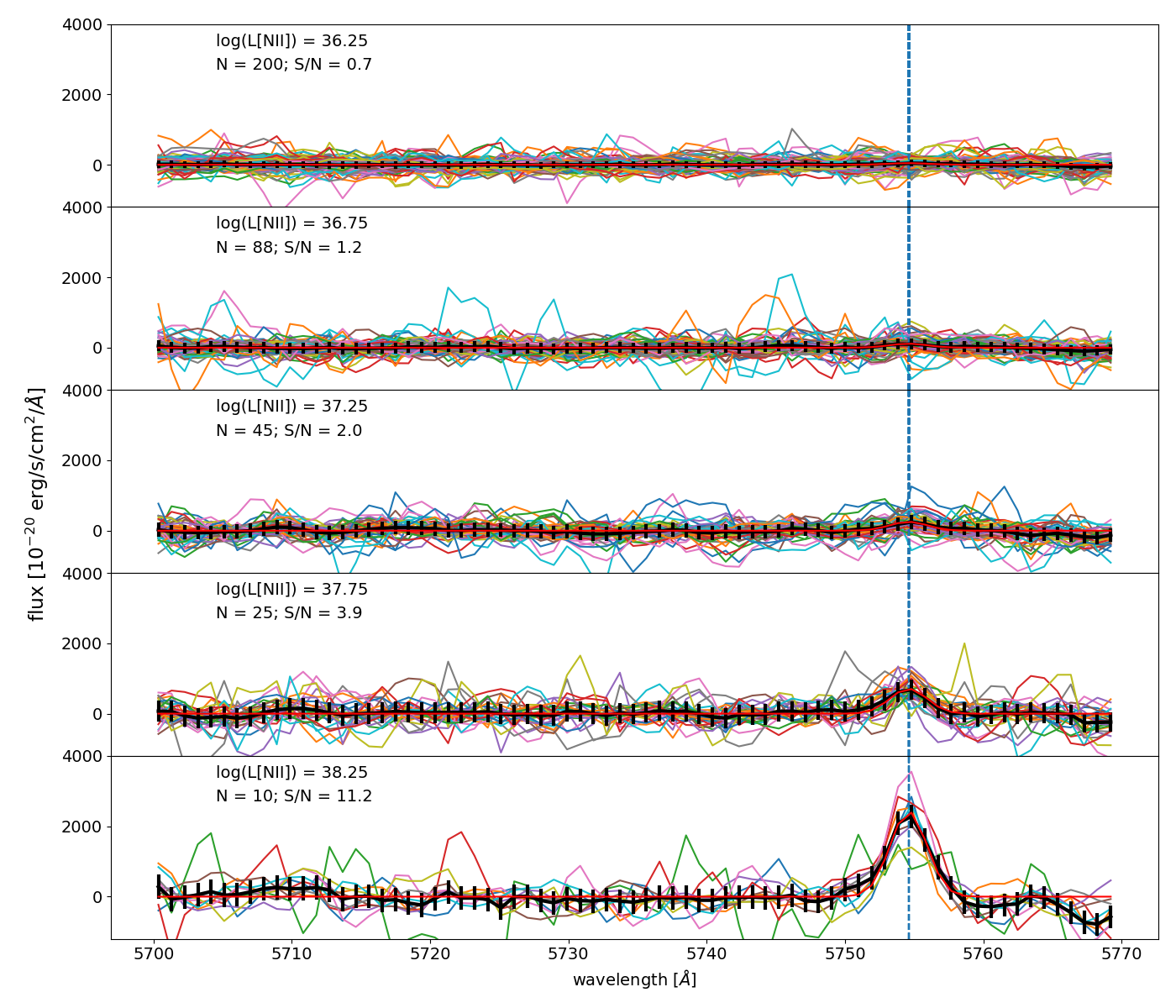}
\caption{An example of the resulting stacked spectra in bins of log(L\nii) = 0.5 at a radius of 1 r$_{\rm eff}$ within the galaxy NGC~628, with all spectra shifted to their rest wavelength based on their \ha\ velocity. Individual \hii\ region spectra are shown in color,  the mean stacked spectrum is shown in black, and the fit is shown in red. This wavelength range focuses on the \niiauroral\ line, marked with a blue dashed line. The luminosity bin, number of individual \hii\ regions included, and resulting S/N of the fit are shown in the upper left corner of each plot.  }
\label{appfig:stack}
\end{figure*}

In Section \ref{sec:stacking} we described our technique for stacking \hii\ regions in bins of luminosity and radius to improve our detection of the \niiauroral\ auroral line. 
In Fig. \ref{appfig:stack} we provide examples of our spectral stacks for a series of luminosity bins in NGC~628, at a fixed radius of 1 r$_{\rm eff}$. Even for large samples of 100-200 \hii\ regions, we generally do not recover auroral line detections at L(\nii) $<$ 10$^{37}$ erg s$^{-1}$, however clear improvements are seen in the brighter luminosity bins. 

\section{Emission-line stars}
\label{app:els}

\begin{figure*}
    \includegraphics[width=1\linewidth]{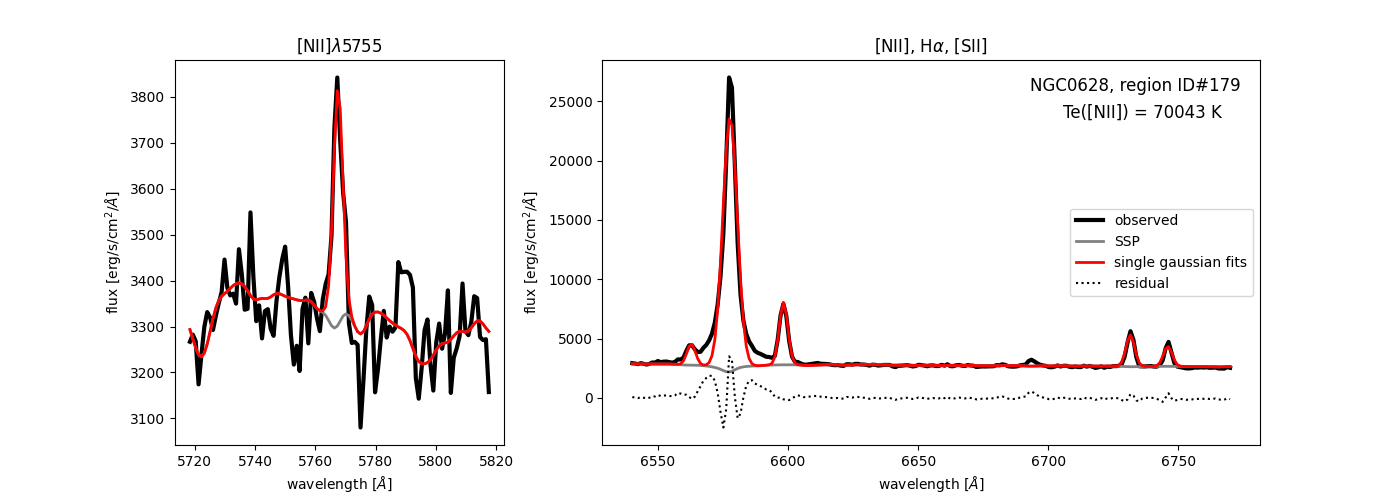}
    \includegraphics[width=1\linewidth]{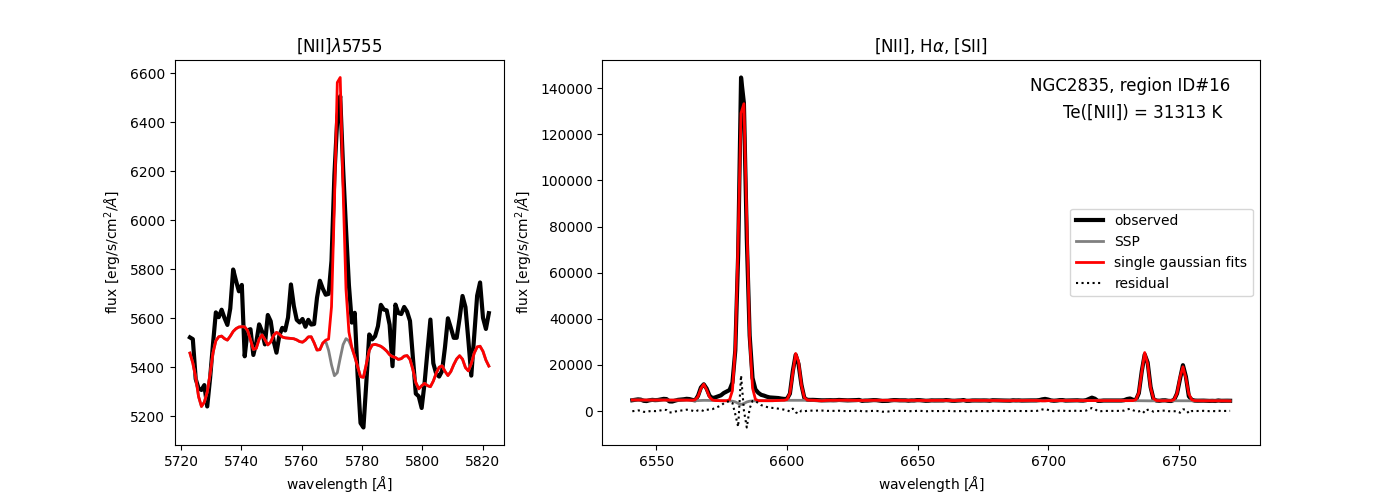}
    \caption{Two extreme outliers identified in our sample, with reported electron temperatures of 70,000 K (top, NGC~628 region ID \#179) and 30,000 K (bottom, NGC~2835 region ID \#16). For each object we show the observed spectrum (black) the SSP model (grey), single Gaussian line fits (red) and fit residuals (dashed). The \niiauroral\ line (left panels) is significantly detected and well fit. Both objects exhibit extremely broad \ha\ wings (right panels). }
    \label{fig:outliers_extreme}
\end{figure*}

In exploring our \te\nii\ catalog \citep{Groves2023} for outliers, we identified two sources (NGC 2835, region ID \#16 and NGC 628, region ID \#179)  with extremely high \te $>$ 20,000~K, large reported errors (above 500 K) but clear \niiauroral\ line detections.   Taken at face value, these objects have \te\ measurements of 70,000~K and 31,000~K. Reducing this to something closer to 10,000~K would require an overestimation of the \niiauroral\ line flux by a factor of five.  
The spectra for the two robust detections, including the SSP model and single Gaussian fits to emission lines, are shown in  Fig. \ref{fig:outliers_extreme}. Interestingly, both have very broad wings in the \ha\ emission, which are not well fit using a single Gaussian, though the \nii$\lambda$6583 line fluxes appear only modestly affected by the blending. 

This unusual broad \ha\ feature makes it likely these objects arise from some type of emission line star, possibly a Wolf-Rayet  star. 
We note that the object identified in NGC~628, region ID\#179 is already reported in \cite{Kreckel2017} and labeled as a possible emission line star. 
In the environments surrounding Wolf-Rayet stars, the wind ejection of enriched material  may result in clumps of nitrogen-enriched material with high densities ($n_{\rm e} > 10,000\,$cm$^{-3}$). If a low electron density is assumed (as would be suggested by the observed integrated \sii\ line ratio), this may result in a subsequent overestimation of the corresponding electron temperature for these sources. 

All other extreme outliers encountered in our catalog are associated with high stellar continuum flux, such that the auroral line detection is likely an artifact of the stellar subtraction procedure.

\section{Alternate strong-line calibrations}
\label{app:altpresc}

In Section \ref{sec:gradients} and Fig. \ref{fig:gradients_scal} we show a comparison of the strong-line S-calibration metallicities with our \te-based measurements. 
Here, we repeat this for alternative metallicity prescriptions, to facilitate a qualitative comparison of the behavior of each compared to the \te-based metallicities. 

\begin{figure*}
\centering
\includegraphics[width=7in]{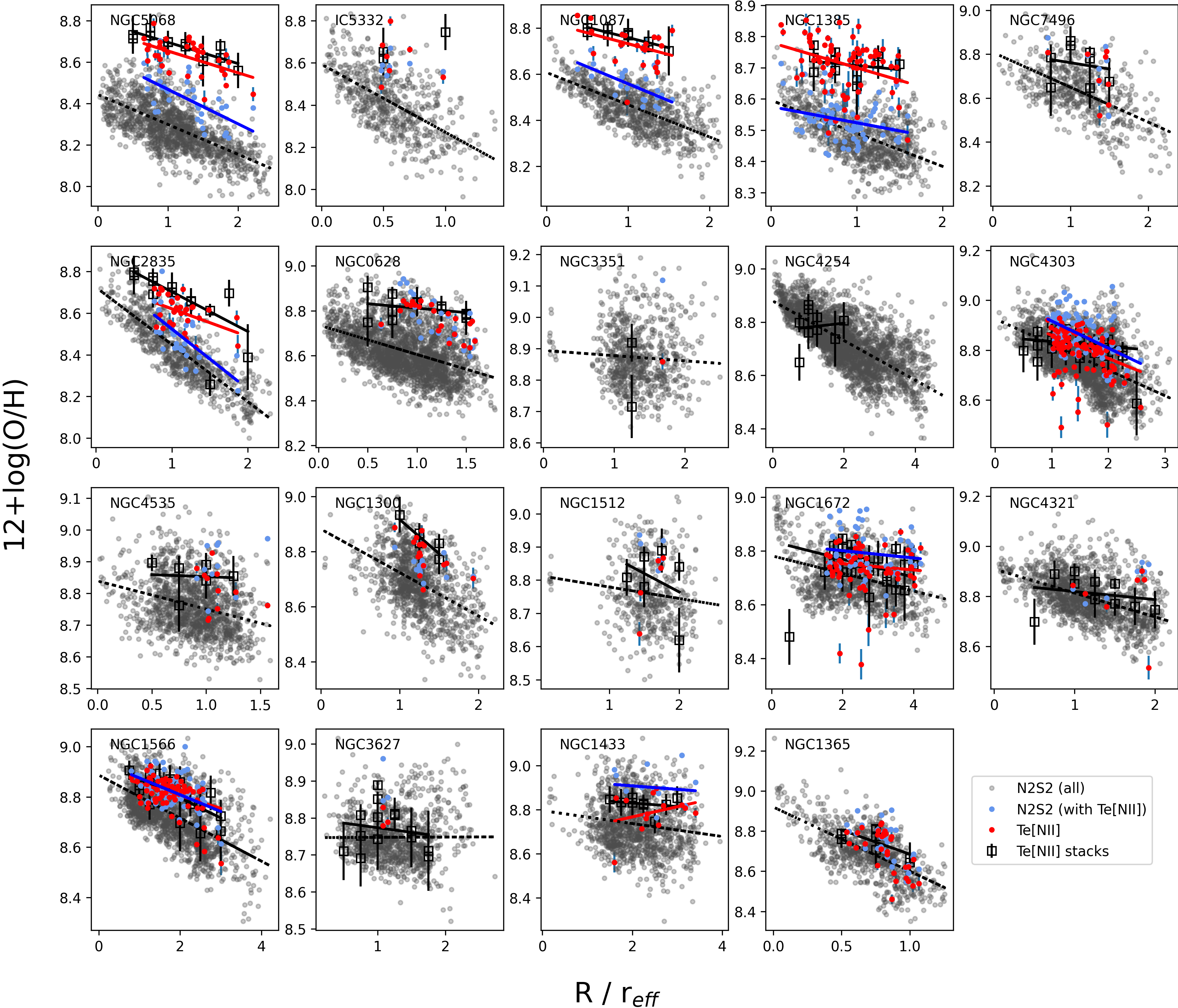}
\caption{A comparison of the our \te-based metallicities with strong-line metallicities in our galaxy sample, as a function of radius. Here, we use the N2S2 \cite{Dopita2016} strong-line calibration. 
The N2S2 values (grey) are compared with individual \te\nii\ metallicities (red) and \te\nii\ stacked metallicities (black). For context, the subsample of individual regions with \te\nii\ detections are highlighted within the N2S2 measurements in light blue. Linear radial gradients are fit when there are at least 5 measurements that cover at least 0.5~r$_{\rm eff}$. Galaxies are ordered from low (top) to high (bottom) stellar mass, and axis scalings are not  matched between galaxies.  
}
\label{appfig:n2s2}
\end{figure*}

\begin{figure*}
\centering
\includegraphics[width=7in]{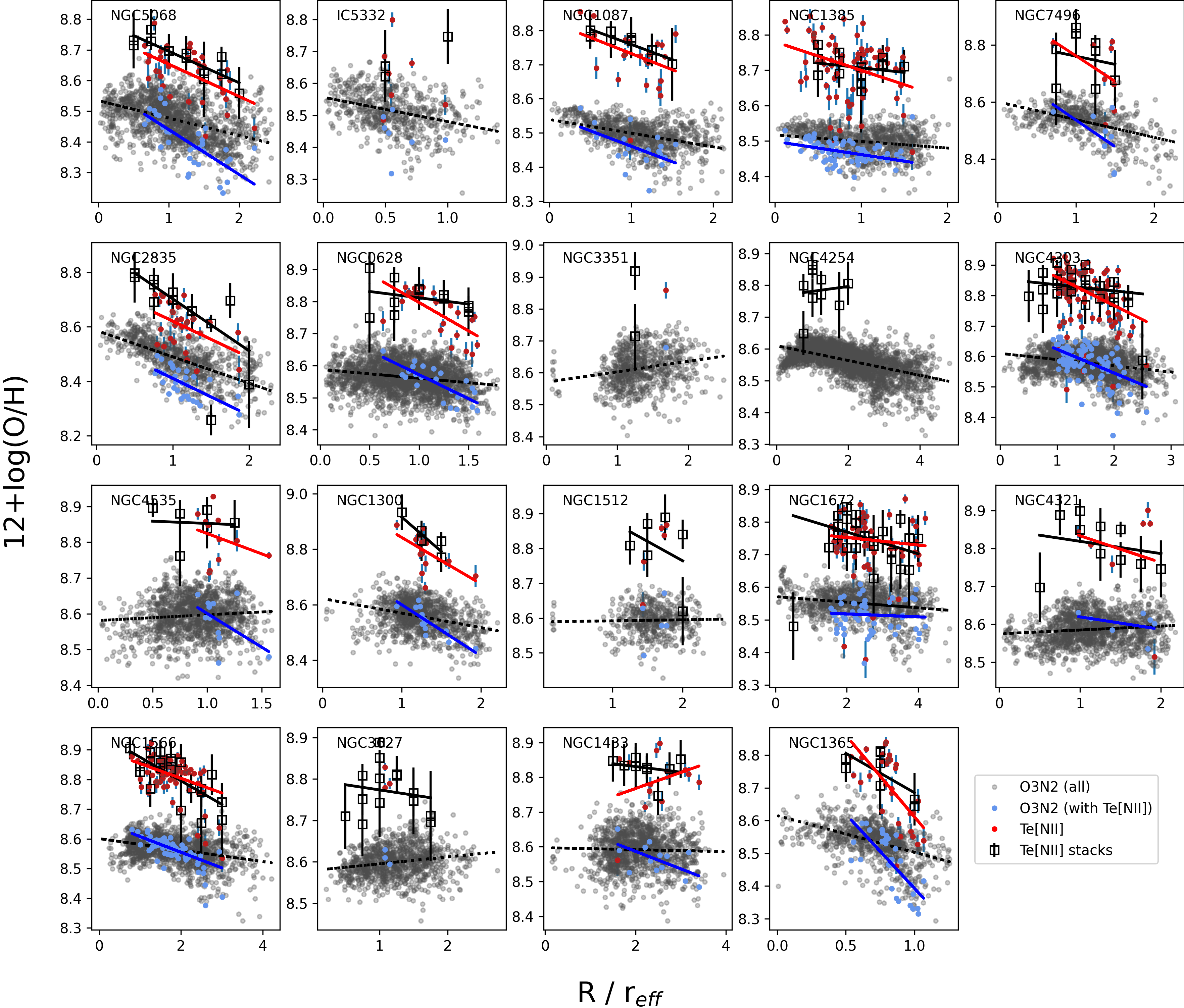}
\caption{
A comparison of the our \te-based metallicities with strong-line metallicities in our galaxy sample, as a function of radius. Here, we use the O3N2 \cite{Marino2013} strong-line calibration. 
The O3N2 values (grey) are compared with individual \te\nii\ metallicities (red) and \te\nii\ stacked metallicities (black). For context, the subsample of individual regions with \te\nii\ detections are highlighted within the O3N2 measurements in light blue. Linear radial gradients are fit when there are at least 5 measurements that cover at least 0.5~r$_{\rm eff}$. Galaxies are ordered from low (top) to high (bottom) stellar mass, and axis scalings are not  matched  between galaxies.  
}
\label{appfig:o3n2}
\end{figure*}

\begin{figure*}
\centering
\includegraphics[width=7in]{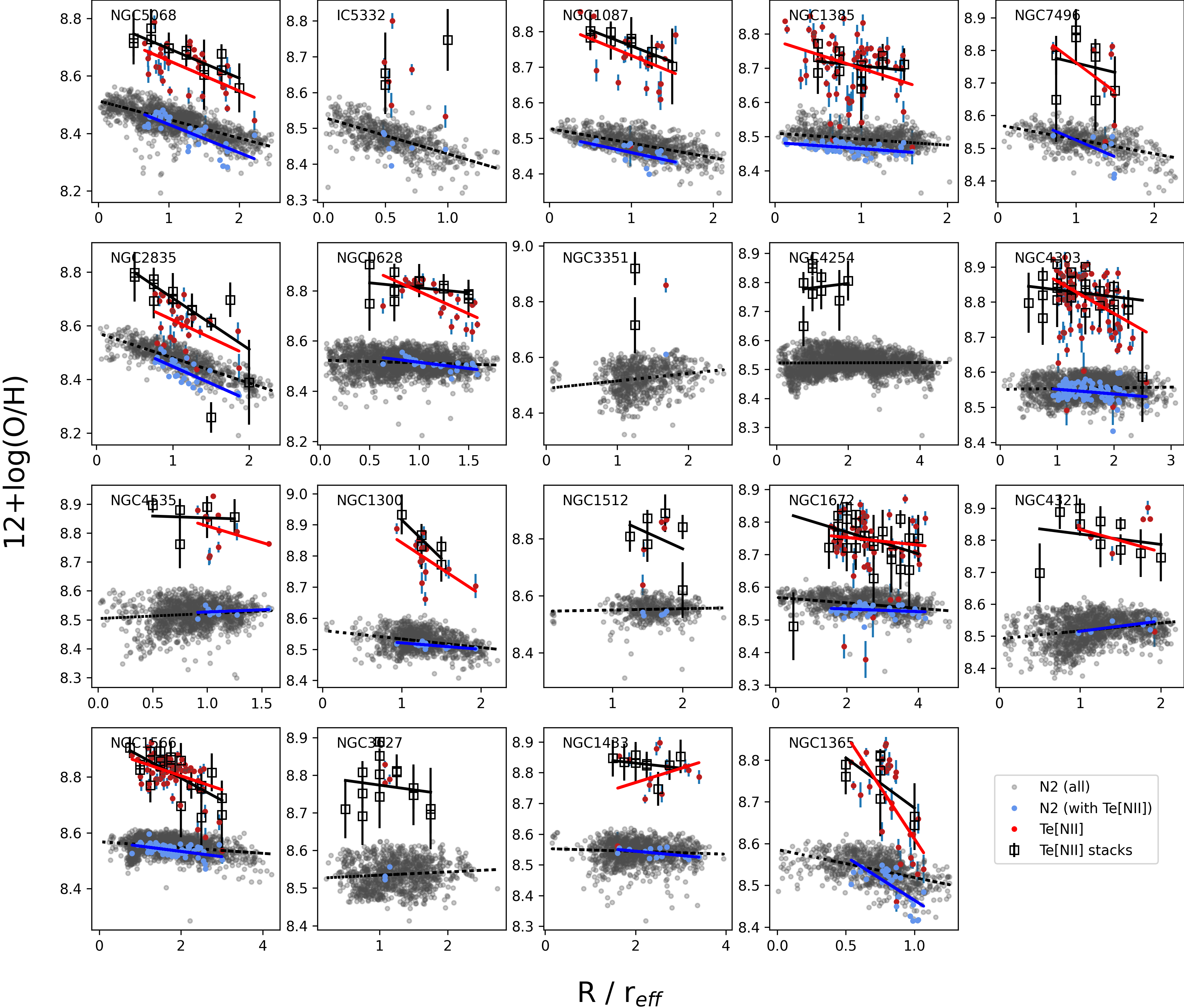}
\caption{A comparison of the our \te-based metallicities with strong-line metallicities in our galaxy sample, as a function of radius. Here, we use the N2 \cite{Marino2013} strong-line calibration. 
The N2 values (grey) are compared with individual \te\nii\ metallicities (red) and \te\nii\ stacked metallicities (black). For context, the subsample of individual regions with \te\nii\ detections are highlighted within the N2 measurements in light blue. Linear radial gradients are fit when there are at least 5 measurements that cover at least 0.5~r$_{\rm eff}$. Galaxies are ordered from low (top) to high (bottom) stellar mass, and axis scalings are not  matched between galaxies.  }
\label{appfig:n2}
\end{figure*}

Figure \ref{appfig:n2s2} uses the \cite{Dopita2016} calibration based on N2S2. 
Figure \ref{appfig:o3n2} uses the \cite{Marino2013} calibration of the O3N2 diagnostic, and Fig. \ref{appfig:n2} uses their N2 diagnostic calibration. 

\begin{figure*}
\centering
\includegraphics[width=7in]{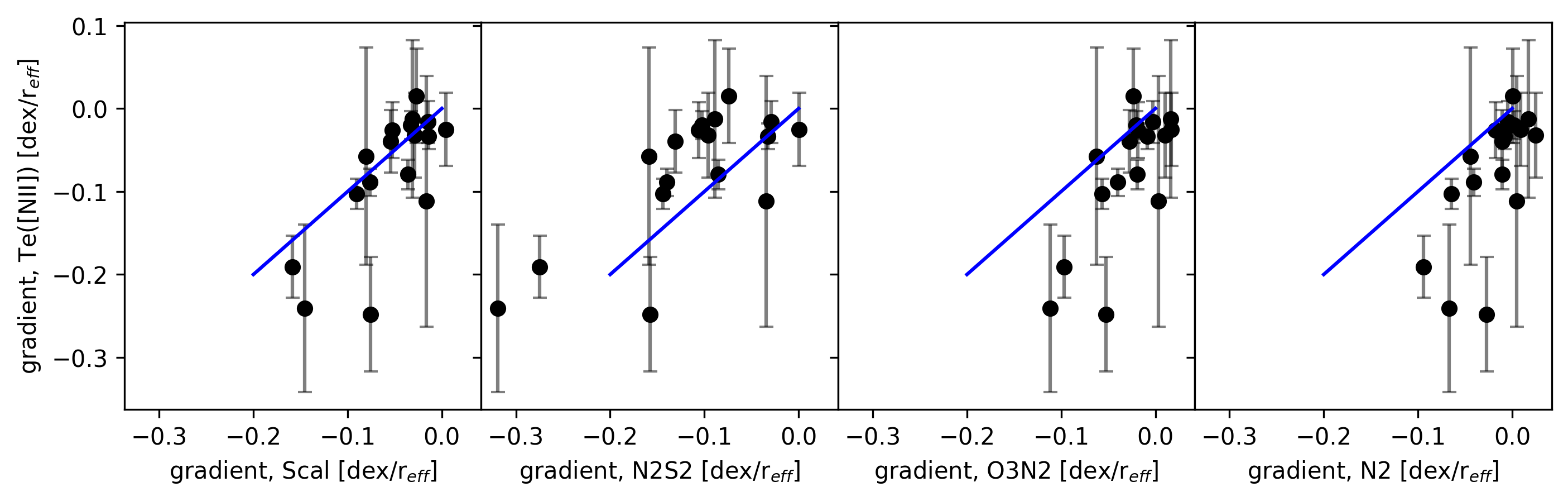}
\caption{A comparison of the metallicity gradient slopes from  our \te-based metallicities with different strong-line calibrations, including the Scal (left), N2S2 (center left), O3N2 (center right), and N2 (right). The blue one-to-one line is shown for reference.}
\label{appfig:compare_gradients}
\end{figure*}

A direct comparison of the slopes derived for each calibration compared to the \te\nii\ metallicity gradients is shown in Figure \ref{appfig:compare_gradients}. While all show clear correlations, O3N2 and N2 show a smaller dynamic range in metallicity gradients. 

\end{appendix}

\end{document}